\documentclass[aps,pra,eqsecnum,amsmath,onecolumn,notitlepage,nofootinbib,superscriptaddress,10pt]{revtex4-1}
\usepackage{custom}

\begin{document}

\title{Non-Gaussian fluctuation dynamics in relativistic fluids}

\author{Xin An}
\email{xin.an@ncbj.gov.pl}
\affiliation{National Centre for Nuclear Research, 02-093 Warsaw, Poland}
\affiliation{Department of Physics and Astronomy, University of North Carolina, Chapel Hill, North Carolina 27599, USA}

\author{G\"{o}k\c{c}e Ba\c{s}ar}
\email{gbasar@unc.edu}
\affiliation{Department of Physics and Astronomy, University of North Carolina, Chapel Hill, North Carolina 27599, USA}

\author{Mikhail Stephanov}
\email{misha@uic.edu}
\affiliation{Department of Physics, University of Illinois, Chicago, Illinois 60607, USA}

\author{Ho-Ung Yee}
\email{hyee@uic.edu}
\affiliation{Department of Physics, University of Illinois, Chicago, Illinois 60607, USA}

\date{\today}

\newpage

\begin{abstract}

  We consider non-equilibrium evolution of {\em non-Gaussian\/}
  fluctuations within relativistic hydrodynamics relevant for the QCD
  critical point search in heavy-ion collision experiments. We rely on
  the hierarchy of relaxation time scales, which emerges in the
  hydrodynamic regime near the critical point, to focus on the slowest
  mode such as the fluctuations of specific entropy, whose equilibrium
  magnitude, non-Gaussianity and typical relaxation time are
  increasing as the critical point is approached.  We derive evolution
  equations for the non-Gaussian correlators of this diffusive mode in
  an arbitrary relativistic hydrodynamic flow. We compare with the
  simpler case of the stochastic diffusion on a static homogeneous
  background and identify terms which are specific to the case of the
  full hydrodynamics with pressure fluctuations and flow.

\end{abstract}

\maketitle

\section{Introduction}
\label{sec:intro}

The physics of thermal fluctuations~\cite{Landau:2013stat2} in
hydrodynamics \cite{Landau:2013fluid} has received renewed interest
recently in the context of relativistic heavy-ion collision
experiments.  Hydrodynamics proved to be remarkably successful in describing the
data from such collisions \cite{Jeon:2015dfa,Romatschke:2017ejr}. One
of the major goals of heavy-ion collision experiments is the discovery
of the QCD critical point -- the end point of the conjectured
first-order transition separating hadron gas and the quark-gluon
plasma phases \cite{Aggarwal:2010cw}. This point is characterized by
a certain singular behavior of fluctuations --
a universal feature of the critical points
\cite{Stephanov:1998dy,Stephanov:1999zu,Stephanov:2004wx}. Therefore,
understanding of fluctuations and, in particular, of their
non-equilibrium evolution in the environment of the hydrodynamically
expanding QCD fireball is important to enable the experimental search
for the QCD critical point~\cite{Bzdak:2019pkr,An_2022}.

The parameter controlling the importance of fluctuations at
wavelengths of order $\ell$ is the ratio of the ``correlation
volume'' of the size of the correlation length $\xi$ to the
``homogeneity volume'' of order $\ell^3$. We shall
generically refer to this parameter as $\varepsilon$. The central limit theorem
suppresses the magnitude of fluctuations as well as their
non-Gaussianity by a power of $\varepsilon$. In a typical condensed matter
experiment $\varepsilon$ is extremely small and thus fluctuations
often play a negligible role. In heavy-ion collisions, where the
system size ($\mathcal O(10 \text{ fm})$) is large compared to the
typical correlation length $\xi$ (a fraction of fm), the
parameter $\varepsilon$ is small but not negligible. Fluctuations are
observable in heavy-ion collisions and play important role in
understanding the thermodynamic properties of the QCD fireball.

Furthermore, near the critical point, as the correlation length
becomes longer, the fluctuations grow and play even more important role.
Non-Gaussianity of fluctuations, also controlled by $\varepsilon$, is
increasing even faster at the critical point than their magnitude
\cite{Stephanov:2008qz}, and is expected to show a specific
nonmonotonic dependence on the collision energy
\cite{Stephanov:2011pb}, which in heavy-ion collisions controls the
thermodynamic conditions at freeze-out. Therefore, non-Gaussian
fluctuation measures have emerged as observables of primary experimental
interest in the search for the QCD critical point \cite{STAR:2021iop}.

Another effect of the critical point is to slow down the relaxation
towards local thermodynamic equilibrium. The critical slowing down
makes it increasingly important to consider non-equilibrium evolution
of fluctuations near the critical point, as earlier estimates and
model calculations
demonstrate~\cite{Stephanov:1999zu,Berdnikov:1999ph,Mukherjee:2015swa}. To
make quantitatively reliable predictions, however, one needs to consider
the non-equilibrium behavior of fluctuations in the ``first
principle'' hydrodynamic approach.

There has been considerable recent effort and progress towards
understanding non-equilibrium evolution of fluctuations in the context
of the hydrodynamics of the heavy-ion collisions with the aim of
mapping the QCD phase diagram (see Ref.~\cite{An_2022} and references
therein). The focus of this paper is on
the approach where fluctuations are described in terms of the
correlation functions obeying deterministic evolution equations, i.e.,
the so-called deterministic (also known as hydrokinetic) approach. This
approach was
introduced and developed recently in the context of relativistic
heavy-ion collisions \cite{Akamatsu:2017,Akamatsu:2018,Martinez:2018},
more generally in
\cite{An:2019rhf,An:2019fdc,An:2020jjk,An_2021,An:2022tfk}, and much earlier, in
the non-relativistic condensed matter context, in
\cite{andreev1970twoliquid,Andreev:1978}.

Within the deterministic approach most of the work has been done on
Gaussian fluctuation measures: two-point correlators, or their
Wigner transforms. In this approach, the first step towards the study
of the {\em non-Gaussian\/} hydrodynamic fluctuations out of
equilibrium was made in Ref.~\cite{An_2021},\footnote{Evolution of
  non-Gaussian {\em cumulants}, i.e., spatial integrals of the
  correlation functions we consider here, was studied previously in
  Refs.~\cite{Mukherjee:2015swa,Mukherjee:2016kyu} and also within the
  complementary stochastic approach, e.g., in
  Ref.~\cite{Nahrgang:2018afz}.  Non-Gaussian correlation functions
  introduced in Ref.~\cite{An_2021} were also studied in the effective
  theory approach in Ref.~\cite{Sogabe:2021svv}. } where the
generalized $n$-point Wigner transform was introduced and the
equations for the corresponding $n$-point Wigner functions were
derived.  Ref.~\cite{An_2021} considered the simplest hydrodynamic
system: nonlinear charge diffusion at constant temperature. Such a
system is characterized by a single hydrodynamic field: conserved
charge density.  Full hydrodynamics necessary to describe heavy-ion
collisions involves (at least) five hydrodynamic fields: conserved
densities of energy, baryon charge and three-momentum of the
fluid. The full theory of hydrodynamic fluctuations is an ambitious
goal and in this paper we shall present another step towards it by
considering fluctuations of both energy and baryon density in the
regime relevant to the critical point search, where the correlation
length becomes large compared to typical microscopic scales (while
still being smaller than $\ell$) and fluctuations of a certain
critical mode dominate.

In this regime an important hierarchy of scales emerges. The slowest
mode is the diffusive (nonpropagating) mode $m=s/n$, i.e., entropy to
baryon number ratio, which we shall refer to as specific entropy. This mode
does not mix with propagating sound oscillations, which makes it
purely diffusive (unlike fluctuations of energy and baryon densities
on their own). Furthermore, unlike the case of other diffusive modes, such as
transverse momentum densities, the diffusion constant for $m$ vanishes
at the critical point making $m$ the slowest mode in this regime. Such
a hierarchy of scales was exploited in the construction of Hydro+ theory
in Ref.~\cite{Stephanov:2018hydro+} as well as in the Hydro++
theory in Ref.~\cite{An:2019fdc}, where the next-to-slowest momentum
diffusion modes were also included. In both cases only two-point
correlators were considered. The goal of this work is to extend the Hydro+
formalism to the non-Gaussian fluctuation measures, i.e., $n$-point
correlation functions.

As in Ref.~\cite{Stephanov:2018hydro+} we shall focus on
fluctuations of $m$, which are important for two
related reasons. First, as explained above, this mode and its
fluctuations are slowest to relax, and therefore are most in need of
nonequilibrium description. Second, the magnitude and
non-Gaussianity measures of the fluctuations of
this mode diverge with the correlation length most strongly (with the
largest critical exponent), thus providing the most sensitive
observable signatures of the critical point.

Furthermore, in Ref.~\cite{An_2021} we considered
fluctuations in a stationary fluid without any flow. Here we shall
treat the most general relativistic flow in a fully Lorentz covariant
formalism necessary for applications to heavy-ion collisions.
We review the formalism which was introduced in
Ref.~\cite{An:2019rhf,An:2019fdc} for two-point correlators with a specific focus
on generalizing it to non-Gaussian fluctuations.
We use the $n$-point
wave number dependent correlation functions which we introduced in
Ref.~\cite{An_2021} by generalizing the well-known Wigner transform and
show how the confluent formalism of Refs.~\cite{An:2019rhf,An:2019fdc} naturally
extends to these objects in
Section~\ref{sec:confl-form-multi}.

While the simple diffusion problem in Ref.~\cite{An_2021}
  contains only one fluctuating field variable (charge density~$n$), we find
that---even in the regime where the fluctuations of $m$ are the slowest
and the fluctuations of the faster hydrodynamic modes, such as
pressure $p$, can be
considered as equilibrated---we cannot neglect these fluctuations, as
they modify equations for the evolution of fluctuations of $m$ via
nonlinearities in the equation of state, or mode coupling.
We discuss this nontrivial part
of the derivation in Section~\ref{sec:entropy_equations}.

We should point out again that we set up the formalism in the most
general form necessary to tackle the full system of hydrodynamic
equations with fluctuations of all hydrodynamic variables. Our
focus on the slowest variable $m$ allows us to make the logical first step
towards considering the full system. This
focus simplifies the calculations and allows us to hone the tools
needed to tackle this ambitious goal.

We shall use the approach of Refs.~\cite{An:2019rhf,An:2019fdc}, i.e.,
expand the stochastic hydrodynamic equations in powers of fluctuation
magnitude. Unlike Refs.~\cite{An:2019rhf,An:2019fdc} we shall not
consider feedback of fluctuations and assume, based on the analysis of
these contributions in Refs.~\cite{An:2019rhf,An:2019fdc}, that the
major effect of these (UV sensitive) contributions have been absorbed
into renormalization of hydrodynamic variables, equation of state and
transport coefficients which, therefore, take physical (cutoff
independent) values. In the diagrammatic representation we introduced
in Ref.~\cite{An_2021} such feedback contributions are represented by
loop diagrams. As in Ref.~\cite{An_2021}, we shall consider only
``tree-level'' terms in the evolution equations for non-Gaussian
correlators. Again, this is not a limitation of the approach, but a
natural simplifying first step in its development.

In Sec.~\ref{sec:general} we introduce the general formalism for
the dynamical evolution equations for the $n$-point correlation
functions in the presence of background hydrodynamic flow. In Sec.~\ref{sec:entropy} we focus on the specific entropy fluctuations for
reasons explained above and, by using the results developed in Sec.~\ref{sec:general}, we derive the evolution equations for the two-,
three- and four-point functions of the specific entropy
fluctuations. Thus we arrive at the main results of this paper which
can be found in Sec.~\ref{sec:entropy_equations}.  We summarize our
findings and discuss the outlook for future developments in Sec.~\ref{sec:conclusions}. 

\section{General deterministic formalism for field fluctuations}
\label{sec:general}

In this section we extend the formalism for the dynamical evolution
equation of relativistic hydrodynamic fluctuations we developed in
Refs.~\cite{An:2019rhf,An:2019fdc} to $n$-point correlation
functions. In particular, we show that the confluent formalism, which
allows us to describe fluctuations in the local rest frame of the
fluid in a natural and fully Lorentz covariant way, generalizes naturally to
$n$-point functions. We also use the multipoint Wigner transform,
which we introduced in Ref.~\cite{An_2021}, to express the $n$-point
correlators in terms of one spatial coordinate of the midpoint and
$n-1$ independent wave vectors. The results of
this section are general, i.e., not limited to a particular
fluctuating variable of variables.

\subsection{Evolution of correlation functions}

Let us first derive the evolution equation for
the $n$-point correlation functions of a set of generic stochastic
field variables $\sth\psi_i$ where the subscript $i$ labels
different local hydrodynamic fields (such as entropy per baryon $m\equiv
s/n$, pressure $p$, and fluid velocity $u$ as in
Refs.~\cite{An:2019rhf,An:2019fdc}). Each of those variables satisfies
its own Langevin-type equation that can be generically written in a
covariant form,
\begin{align}\label{eq:dtpsi}
  \sth u\cdot\partial\sth\psi_i=\sth F_i+\xi_i\,,
  \quad\mbox{with}\quad
  \xi_i=\sth H_{ij}\eta_j\,,
\end{align}
where the symbol ``$~\sth~~$'' denotes a stochastic quantity. Functions or
functionals, such as $\sth u\equiv u(\sth\psi_i)$ or
$\sth F\equiv F(\sth\psi)$  inherit
the stochastic symbol ``$~\sth~~$'' from their arguments. The four-velocity
obeys $\sth u^2=-1$, and should be understood as a four-vector function
of $\psi_i$'s (or can also be chosen to be among the variables $\psi_i$
as in Refs.~\cite{An:2019rhf,An:2019fdc}). $\sth F_i$ is the drift
``force'' and 
$\xi_i$ is the noise
(random ``force'') for the variable $\psi_i$, expressed in terms of the
canonically normalized local Gaussian noise\footnote{As discussed in
  Ref.~\cite{An_2021}, in hydrodynamics, the contribution of the
  non-Gaussianity of the noise will not appear in the leading order in
  the hydrodynamic (i.e., gradient) expansion.}:
\begin{align}\label{eq:eta-eta}
\av{\eta_i}=0\,,\qquad	\av{\eta_{i}(x_1)\eta_{j}(x_2)}=2\delta_{ij}\delta^{(4)}(x_1-x_2)\,.
\end{align}
We include multiplicative noise, since $\sth H = H(\sth\psi)$, and 
define the product $\sth H\eta$ in terms of the It${\bar {\rm o}}$
calculus.\footnote{In practice that means that $H(\sth\psi)$ and $\eta$
  are considered uncorrelated. Under time discretization this
  corresponds to evaluating $H(\psi(t))$ and $\eta(t)$ at the same
  time point $t$. Different discretization prescriptions, such as
  Stratonovich, where $H=H((\psi(t) + \psi(t+\Delta t))/2)$, can be used
  to describe the same physics with a given equilibrium distribution
  of fluctuating variables, as long as the drift term is chosen
  accordingly. Below we shall verify a posteriori that our equations
  reproduce the correct equilibrium values of the fluctuations given
  by thermodynamics. This provides a check of the consistency of the
  implementation of the Ito prescription in our approach.}

The
Onsager matrix (operator)  $Q=Q^T$ is the ``square'' of $H$:
$Q_{ij}\equiv H_{ik}H_{jk}$.

Denoting fluctuation of a given stochastic quantity $\sth X$ as
\begin{equation}
  \label{eq::X:}
  \fl{\sth X}\equiv \delta X\equiv \sth X - \langle \sth X \rangle \equiv \sth X - X\,,
\end{equation}
and introducing the fluctuation of the variables $\psi_i$,
\begin{align}\label{eq:phi-def}
	\phi_i\equiv\,:\!\spsi_i\!:\,\equiv\spsi_i-\psi_i\,,
\end{align}
we can expand the stochastic
equations~(\ref{eq:dtpsi}) in powers of $\phi_i$, using, e.g.,
\begin{align}\label{eq:u-expansion}
	\sth F_i=\sum_{n=0}^\infty\frac{1}{n!}F_{i,\,j_1\dots
  j_n}\phi_{j_1}\dots\phi_{j_n}=F_i+F_{i,\,j_1}\phi_{j_1}
  +\frac12 F_{i,\,j_1j_2}\phi_{j_1}\phi_{j_2}+\dots\,,
\end{align}
where $F\equiv F(\psi)$, etc.\ ,
and obtain the evolution equation for fluctuation field $\phi_i$:
\begin{align}\label{eq:dtphi}
	u\cdot\partial\phi_i=\fl{\su\cdot\partial\spsi_i-\du\cdot\partial\spsi_i:\,\,=&:\sum_{n=1}^\infty\frac{1}{n!}L_{i,\,j_1\dots j_n}\phi_{j_1}\dots\phi_{j_n}+\xi_i}\,,
\end{align}
where the term involving $\du$ in Eq.~\eqref{eq:dtphi} must be kept if
the fluctuation of velocity is taken into account and
\begin{equation}\label{eq:L}
	L_{i,\,j_1j_2\dots j_n}=F_{i,\,j_1j_2\dots j_n}-u^\mu_{,\,j_1\dots j_n}(\partial_\mu\psi_i)-\left[n\delta_{ij_1}u^\mu_{,\,j_2\dots j_{n}}\partial_\mu^{(j_1)}\right]_{\overline{1\dots n}}\,.
      \end{equation}
 Here $\partial_\mu^{(j_1)}$ acts on $\phi_{j_1}$ only, and the Einstein summation rule over repeated indices is implied
 throughout this paper. The
subscript $\overline{1\dots n}$ denotes the ``averaging over
permutations'', i.e., the sum over all $n!$ permutations of all
composite index-position labels $(j_1,x_1),\dots,(j_n,x_n)$ divided by
$n!$,
  \begin{equation}\label{eq:aveperm}
\Big[\dots\Big]_{\overline{1\dots n}}=\frac1{n!}\Big[\dots\Big]_{{\rm P}_{1\dots n}}\,,
\end{equation}
where we used the notation of Ref.~\cite{An_2021}, ${{\rm P}_{1\dots
    n}}$, to denote the sum over permutations. Furthermore, since Eqs.~(\ref{eq:dtpsi}) that
 we consider are {\em differential} equations,  each
 expansion coefficient such as $L_{i,\,j_1j_2\dots j_n}$ must be treated as
 a multilinear 
 differential operator.
  These operators are linear in
 each of their arguments, which are identified by the matching
 repeated indices. It is
 helpful, in this regard, to think of the arguments $x_i$ of the field
 $\phi_i(x_i)$ as a part of the composite label $(i,x_i)$ denoting
 position as well as the name or index of the variable. 
 
Having this at hand it is
straightforward to write down the evolution equation for
the ``raw'' $n$-point correlation functions
\begin{equation}\label{eq:Gn_def}
  G_{i_1\dots i_n}\equiv\av{\phi_{i_1}(x_1)\dots\phi_{i_n}(x_n)}
\end{equation}
by taking the time derivative
in the rest frame of the fluid at midpoint $x\equiv \sum_{i=1}^n
x_i/n$:
\begin{multline}\label{eq:dtGn_full}
  u\cdot\partial^{(x)} G_{i_1\dots i_n}\equiv\sum_{i=1}^n u\cdot\frac{\partial}{\partial x_i} G_{i_1\dots i_n}=n\left[u\cdot\frac{\partial}{\partial x_1} G_{i_1\dots i_n}\right]_{\overline{1\dots n}}=n\Bigg\{-\left(y_1\cdot\partial u\right)\cdot\frac{\partial}{\partial x_1}G_{i_1\dots i_n}\\[1pt]
   + \sum_{m=0}^\infty
    \frac{1}{m!}\Big[L_{i_1,\,j_1\dots j_m}(G_{j_1\dots j_mi_2\dots i_n}-G_{j_1\dots j_m}G_{i_2\dots i_n})+(n-1)Q_{i_1i_2,\,j_1\dots j_m}G_{j_1\dots j_mi_3\dots i_n}\Big]\Bigg\}_{\overline{1\dots n}}\,,
\end{multline}
where we have used\footnote{Eq.~(\ref{eq:uddx1}) assumes that the
  velocity $u$ varies slowly on the characteristic scale of the
  correlations we study, as is the case in hydrodynamics. This
  property of hydrodynamics underlines the approach known as
  ``hydrokinetics''
  \cite{Akamatsu:2017,Akamatsu:2018,Martinez:2018,An:2019rhf,An:2019fdc}. Eqs.~(\ref{eq:dtGn_full})--(\ref{eq:uddx1-2}), for $n=2$ and truncated at
  $m=1$, upon Wigner transform reproduce the equations for Gaussian
  fluctuation correlators known as ``hydrokinetic equations". Our
  equations are more general and include non-Gaussian fluctuations and
multiplicative noise.}
\begin{align}\label{eq:uddx1}
  &u\cdot\frac{\partial}{\partial x_1} G_{i_1\dots i_n}= \left[u(x_1)-\left(y_1\cdot\partial u\right)\right]\cdot\frac{\partial}{\partial x_1}G_{i_1\dots i_n}
\end{align}
and
\begin{align}\label{eq:uddx1-2}
  &u(x_1)\cdot\frac{\partial}{\partial x_1} G_{i_1\dots i_n}= u(x_1)\cdot\frac{\partial}{\partial x_1}\av{\phi_{i_1}(x_1)\dots\phi_{i_n}(x_n)}\nn
  =& \sum_{m=0}^\infty
    \frac{1}{m!}\av{:\!L_{i_1,\,j_1\dots j_m}(x_1)\phi_{j_1}(x_1)\dots\phi_{j_m}(x_1)\!:\phi_{i_2}(x_2)\dots\phi_{i_n}(x_n)}\nn
  &+\frac{n-1}{m!}\av{Q_{i_1i_2,\,j_1\dots j_m}(x_1,x_2)\phi_{j_1}(x_1)\dots\phi_{j_m}(x_1)\phi_{i_3}(x_3)\dots\phi_{i_n}(x_n)}_{\overline{2\dots n}}\nn
  =& \sum_{m=0}^\infty\frac{1}{m!}\Big[L_{i_1,\,j_1\dots j_m}(G_{j_1\dots j_mi_2\dots i_n}-G_{j_1\dots j_m}G_{i_2\dots i_n})+(n-1)\left(Q_{i_1i_2,\,j_1\dots j_m}G_{j_1\dots j_mi_3\dots i_n}\right)_{\overline{2\dots n}}\Big]\,.
\end{align}

By the first equality in Eq.~\eqref{eq:dtGn_full} we defined the
differential operator, which can be understood as the derivative with
respect to the midpoint $x$ (see also Eq.~\eqref{eq:d/dx-d/dy}). The
subscripts $\overline{1\dots n}$ in Eq.~\eqref{eq:dtGn_full},
and $\overline{2\dots n}$ in Eq.~(\ref{eq:uddx1}), denote the ``averaging over
permutations'' defined in Eq.~\eqref{eq:aveperm}. The factors
$n=n!/(n-1)!$ and $n(n-1)=n!/(n-2)!$ in 
Eq.~\eqref{eq:dtGn_full} conveniently match the number of the
nonidentical terms among the $n!$ permutations. The identical terms
arise due to the invariance of the correlators in
Eq.~(\ref{eq:Gn_def}) under the
permutation ${\rm P}_{1\dots n}$. The function $Q(x_1,x_2)$ is the integral kernel of the Onsager
operator, i.e., $Q(x_1,x_2)\equiv Q \delta^{(3)}(x_1-x_2)$, 
where the
delta-function support is a line parallel to $u(x)$, i.e.,
$\delta^{(3)}(y)=0$ unless $u(x)\cdot y=0$. The first term on the right-hand side of Eq.~\eqref{eq:dtGn_full}
arises because the velocity at midpoint $u(x)$ on the
left-hand side is different from the velocity $u(x_1)$ involved in the
equation of motion for $\phi(x_1)$ by the amount proportional to
$y_1$.

\subsection{Power counting and perturbation theory}
\label{sec:power-count-pert}

Evolution equations~\eqref{eq:dtGn_full} form an infinite system of
equations for an infinite set of correlation functions. However, in the regime
of applicability of hydrodynamics, there is a natural hierarchy in
this system which allows us to systematically truncate
and solve these equations. In this section we discuss this hierarchy
and the small power counting parameter(s) which control it (see also Refs.~\cite{An:2019fdc,An_2021}).

Hydrodynamics is an effective theory that emerges when the thermal state
of the system is sufficiently homogeneous. More precisely,
hydrodynamic variables (i.e., conserved densities) characterizing the
local thermal state of the system must vary on a scale $\ell$ much
longer than the microscopic scale $\ell_{\rm mic}$, which typically is
of order of the correlation length $\xi$.\footnote{The vicinity
of the critical point is characterized by the correlation length $\xi$
becoming much larger than other microscopic scales. In this
interesting case we still require $\ell\gg\xi$, a reasonable condition
for the scales relevant for the heavy-ion collisions.
Dynamic critical behavior in the
regime $\ell\sim\xi$, which is beyond the scope of this work, can be described
by a hydrodynamic-like theory such as
model-H~\cite{Hohenberg:1977}. In this regime the power counting we
use begins to break down.}
Indeed, without such a scale separation
one could not describe the state of the system entirely in terms of
thermodynamic (conserved) quantities and, e.g., the thermodynamic
concept of equation of state would lose meaning.  The scale separation
$\ell\gg\ell_{\rm mic}$ is the essential property of the hydrodynamic
regime and it gives rise to a small parameter
$\ell_{\rm mic}/\ell$ (Knudsen number) which controls the
gradient expansion of hydrodynamic equations and ensures their
locality.

When we consider fluctuations in hydrodynamics the relevant
characteristic inhomogeneity scale is the wavelength of the
fluctuations, or the inverse of their wave number $q$, and the
corresponding small parameter is $\varepsilon_q \equiv q\ell_{\rm mic}$.

The small parameter which controls the magnitude and importance of
fluctuations is closely related to $\varepsilon_q$. This parameter is the
inverse of the typical number of the uncorrelated microscopic cells (of volume
$\xi^3$) in the homogeneity region (of volume $\ell^3$, or
$1/q^3$): $\varepsilon\equiv(\xi/\ell)^3\sim (q\xi)^3 $. The central limit theorem guarantees that (1) the magnitude
of the fluctuations averaged over the scale of $\ell\sim1/q$ is suppressed by
$\sqrt\varepsilon$ and (2) the non-Gaussian connected correlators of order $n>2$
are suppressed by $\varepsilon^{n-1}$.

Both $\varepsilon_q$ and $\varepsilon$ are small
in the hydrodynamic regime, when $q\ell_{\rm mic}\ll1$, and $\varepsilon\sim\varepsilon_q^3$. However, for the sake
of generality, and to emphasize the difference between the gradient
expansion corrections and the fluctuation corrections to
hydrodynamics, we shall treat these two parameters as independent.\footnote{There are field theories where these
  two parameters are {\em parametrically} different, e.g., in large
  $N$ theories where
   $\varepsilon\sim1/N$, while $\varepsilon_q\sim N^0$.}

The
corresponding power
counting for various quantities involved in our calculation is as follows:
\begin{equation}\label{eq:powercounting}
  G_{i_1\dots i_n}\sim\varepsilon^{[n/2]}\,,\quad
  G^c_{i_1\dots i_n}\sim\varepsilon^{n-1}\,,
  \quad L_{i_1,\,i_2\dots i_n}\sim \varepsilon_q + \mathcal O(\varepsilon_q^2)\,, \quad
  Q_{i_1i_2,\,j_1\dots j_m}\sim\varepsilon_q^2\varepsilon\,, \quad
  u\cdot\partial\sim\varepsilon_q^2\,,
\end{equation}
where $[n/2]=n/2$ for even $n$ and $(n+1)/2$ for odd $n$, while $G^{c}$ denote
connected correlation functions defined below in Eq.~\eqref{eq:Gn-Gn_c} or \eqref{eq:Gn_c-Gn}. The factor $(\sqrt\varepsilon)^n$ in $G$ is a consequence of the suppression of
the magnitude of fluctuations averaged over hydrodynamic scale by a
factor $\sqrt\varepsilon$. The factor $\varepsilon^{n-1}$ in $G^{\rm
  c}$ results from the requirement that $n-1$ internal correlations
are needed for a fully connected $n$-point correlator (with each internal two-point correlator being of order $\varepsilon$). The factor
$\varepsilon_q$ in~$L$ reflects the fact that the rate of ideal
hydrodynamic evolution for fluctuations is first order in spatial
derivatives, while the  $\mathcal O(\varepsilon_q^2)$ term in~$L$
represents dissipative terms in hydrodynamic equations which are
second order (as in diffusion). The $\varepsilon_q$ term is absent for
purely diffusive variables. The factor $\varepsilon_q^2$ in the
Onsager operator $Q$ controls the magnitude of the noise and is
determined by the fluctuation-dissipation theorem, while the factor
$\varepsilon$ is due to the locality (short range correlation)
of the noise and its suppression on the hydrodynamic scale by the
central limit theorem.

The first few equations (with $n=2,3,4$) of Eq.~\eqref{eq:dtGn_full} are truncated in the double expansion of $\varepsilon$ and $\varepsilon_q$, i.e.,
\begin{subequations}\label{eq:dtGn}
\begin{align}
	u\cdot\partial^{(x)} G_{i_1i_2}(x_1,x_2)
	=&\,2\Big[-\left(y_1\cdot\partial u\right)\cdot\frac{\partial}{\partial x_1}G_{i_1i_2}(x_1,x_2)+L_{i_1,\,j_1}(x_1)G_{j_1i_2}(x_1,x_2)\nn[2pt]
  +&\frac{1}{2}L_{i_1,\,j_1j_2}(x_1)G_{j_1j_2i_2}(x_1,x_1,x_2)+\frac{1}{6}L_{i_1,\,j_1j_2j_3}(x_1)G_{j_1j_2j_3i_2}(x_1,x_1,x_1,x_2)\nn[3pt]
  +&Q_{i_1i_2}(x_1,x_2)+\frac{1}{2}Q_{i_1i_2,\,j_1j_2}(x_1,x_2)G_{j_1j_2}(x_1,x_1)+\mathcal{O}(\varepsilon_q^2\varepsilon^3)\Big]_{\overline{12}}\,,\label{eq:dtG2}
\end{align}
\begin{align}
	u\cdot\partial^{(x)} G_{i_1i_2i_3}(x_1,x_2,x_3)
	=&\,3\Big[-\left(y_1\cdot\partial u\right)\cdot\frac{\partial}{\partial x_1}G_{i_1i_2i_3}(x_1,x_2,x_3)+L_{i_1,\,j_1}(x_1)G_{j_1i_2i_3}(x_1,x_2,x_3)\nn[2pt]
	&+\frac{1}{2}L_{i_1,\,j_1j_2}(x_1)\left[G_{j_1j_2i_2i_3}(x_1,x_1,x_2,x_3)-G_{j_1j_2}(x_1,x_1)G_{i_2i_3}(x_2,x_3)\right]\nn[3pt]
	&+2Q_{i_1i_2,\,j_1}(x_1,x_2)G_{j_1i_3}(x_1,x_3)+\mathcal{O}(\varepsilon_q^2\varepsilon^3)\Big]_{\overline{123}}\,,\label{eq:dtG3}
\end{align}
\begin{align}
	u\cdot\partial^{(x)} G_{i_1i_2i_3i_4}(x_1,x_2,x_3,x_4)
  =&\,4\Big[-\left(y_1\cdot\partial u\right)\cdot\frac{\partial}{\partial x_1}G_{i_1i_2i_3i_4}(x_1,x_2,x_3,x_4)+L_{i_1,\,j_1}(x_1)G_{j_1i_2i_3i_4}(x_1,x_2,x_3,x_4)\nn[2pt]
	&+\frac{1}{2}L_{i_1,\,j_1j_2}(x_1)[G_{j_1j_2i_2i_3i_4}(x_1,x_1,x_2,x_3,x_4)-G_{j_1j_2}(x_1,x_1)G_{i_2i_3i_4}(x_2,x_3,x_4)]\nn[2pt]
	&+\frac{1}{6}L_{i_1,\,j_1j_2j_3}(x_1)G_{j_1j_2j_3i_2i_3i_4}(x_1,x_1,x_1,x_2,x_3,x_4)\nn[5pt]
	&+3Q_{i_1i_2}(x_1,x_2)G_{i_3i_4}(x_3,x_4)+3Q_{i_1i_2,\,j_1}(x_1,x_2)G_{j_1i_3i_4}(x_1,x_3,x_4)\nn[3pt]
	&+\frac{3}{2}Q_{i_1i_2,\,j_1j_2}(x_1,x_2)G_{j_1j_2i_3i_4}(x_1,x_1,x_3,x_4)+\mathcal{O}(\varepsilon_q^2\varepsilon^4)\Big]_{\overline{1234}}\,,\label{eq:dtG4}
\end{align}
\end{subequations}
and equations for $n\geqslant5$ can be obtained accordingly. Here we
only consider the first-order hydrodynamics. Thus the equations for all
multipoint functions are truncated at order
$\varepsilon_q^2$.\footnote{The first term on the right-hand side of
  each equation in \eqref{eq:dtGn}, contains the gradient of the
  background velocity, $\partial u$, and is thus
of order $k$, the characteristic background wave number, which we treat as being of the same order as
$q^2\sim\varepsilon_q^2$, as
in Refs.~\cite{An:2019rhf,An:2019fdc}.} While we keep
only the leading order terms in $\varepsilon$ in Eqs.~\eqref{eq:dtG3} and \eqref{eq:dtG4}, both leading and next-to-leading order terms are kept in Eq.~\eqref{eq:dtG2}, for the purpose of deriving the evolution equation for the four-point connected function, Eq.~\eqref{eq:dtG4_c}.

\begin{figure}[ht]
  \centering  
  \includegraphics[scale=.58]{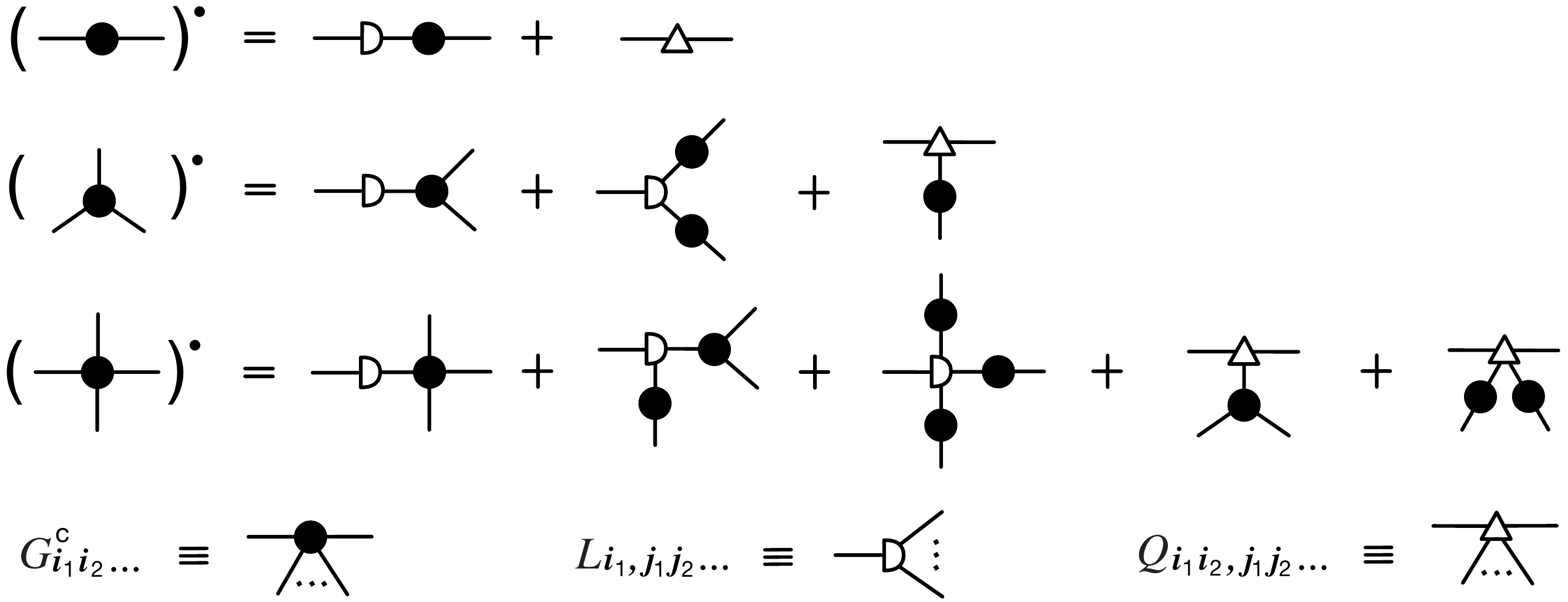}
  \caption{Diagrammatic representation of the evolution
equations~(\ref{eq:dtGn_c}) for multipoint connected correlation
functions.  The dot on the left-hand side of each diagrammatic
equation represents material derivative $u\cdot\partial$.
The open half-circle with two legs represents
$L_{i_1,\,j_1}-\delta_{i_1j_1}(y_1\cdot\partial u)\cdot(\partial/\partial
x_1)$, i.e., it includes the first term on the right-hand side of each
equation in (\ref{eq:dtGn_c}). It is straightforward to generate the
evolution equation for $G^{\rm c}_{n+1}$ recursively by adding
in each diagram representing equation for $G^{\rm c}_n$ either one new leg to
filled vertices or one new $G_2$ correlator to open vertices. This
recursive rule can be readily verified using
Eq.~\eqref{eq:dtGn_c_general}; see also
Fig.~\ref{fig:diagrams-n}. Similar diagrammatic representation applies
to evolution equations~(\ref{eq:LW_c-m}) we derive below.}
  \label{fig:diagrams}
\end{figure}

Our purpose is to derive the evolution equations for connected
correlation functions, which can be directly related to the 
correlations of particles in experiments~\cite{Pradeep:2022eil}.
To this end we would need to
express the correlation functions $G$ in terms of the connected
correlation functions $G^{\rm c}$ (and vice versa). The equations relevant for our calculation of evolution equations up to $n=4$ are given by
\begin{equation}\label{eq:Gn-Gn_c-example}
\begin{gathered}
  G_{i_1i_2}=G^{\rm c}_{i_1i_2}\,, \qquad G_{i_1i_2i_3}=G^{\rm c}_{i_1i_2i_3}\,, \qquad G_{i_1i_2i_3i_4}=\left[G^{\rm c}_{i_1i_2i_3i_4}+3G^{\rm c}_{i_1i_2}G^{\rm c}_{i_3i_4}\right]_{\overline{1234}}\,, \\[7pt]
  G_{i_1i_2i_3i_4i_5}=\left[G^{\rm c}_{i_1i_2i_3i_4i_5}+10G^{\rm c}_{i_1i_2}G^{\rm c}_{i_3i_4i_5}\right]_{\overline{12345}}\,,\\[7pt]
  G_{i_1i_2i_3i_4i_5i_6}=\left[G^{\rm c}_{i_1i_2i_3i_4i_5i_6}+15G^{\rm c}_{i_1i_2}G^{\rm c}_{i_3i_4i_5i_6}+10G^{\rm c}_{i_1i_2i_3}G^{\rm c}_{i_4i_5i_6}+15G^{\rm c}_{i_1i_2}G^{\rm c}_{i_3i_4}G^{\rm c}_{i_5i_6}\right]_{\overline{123456}}\,.
\end{gathered} 
\end{equation}
Using Eq.~\eqref{eq:dtGn} and \eqref{eq:Gn-Gn_c-example} and keeping
the leading order terms in $\varepsilon$, we obtain the
evolution equations for the ``raw'' connected functions for $n=2,3,4$:
\begin{subequations}\label{eq:dtGn_c}
\begin{align}
  &\,u\cdot\partial^{(x)} G^{\rm c}_{i_1i_2}(x_1,x_2)\nn[2pt]
  =&\,2\Big[-\left(y_1\cdot\partial u\right)\cdot\frac{\partial}{\partial x_1}G^{\rm c}_{i_1i_2}(x_1,x_2)+L_{i_1,\,j_1}(x_1)G^{\rm c}_{j_1i_2}(x_1,x_2)+Q_{i_1i_2}(x_1,x_2)+\mathcal{O}(\varepsilon_q^2\varepsilon^2)\Big]_{\overline{12}}\,,\\[10pt]
  &\,u\cdot\partial^{(x)} G^{\rm c}_{i_1i_2i_3}(x_1,x_2,x_3)\nn[2pt]
  =&\,3\Big[-\left(y_1\cdot\partial u\right)\cdot\frac{\partial}{\partial x_1}G^{\rm c}_{i_1i_2i_3}(x_1,x_2,x_3)+L_{i_1,\,j_1}(x_1)G^{\rm c}_{j_1i_2i_3}(x_1,x_2,x_3)\nn[2pt]
  &+L_{i_1,\,j_1j_2}(x_1)G^{\rm c}_{j_1i_2}(x_1,x_2)G^{\rm c}_{j_2i_3}(x_1,x_3)+2Q_{i_1i_2,\,j_1}(x_1,x_2)G^{\rm c}_{j_1i_3}(x_1,x_3)+\mathcal{O}(\varepsilon_q^2\varepsilon^3)\Big]_{\overline{123}}\,,\\[10pt]
  &\,u\cdot\partial^{(x)} G^{\rm c}_{i_1i_2i_3i_4}(x_1,x_2,x_3,x_4)\nn[2pt]
  =&\,4\Big[-\left(y_1\cdot\partial u\right)\cdot\frac{\partial}{\partial x_1}G^{\rm c}_{i_1i_2i_3i_4}(x_1,x_2,x_3,x_4)+L_{i_1,\,j_1}(x_1)G^{\rm c}_{j_1i_2i_3i_4}(x_1,x_2,x_3,x_4)\nn[2pt]
  &+3L_{i_1,\,j_1j_2}(x_1)G^{\rm c}_{j_1i_2}(x_1,x_2)G^{\rm c}_{j_2i_3i_4}(x_1,x_3,x_4)+L_{i_1,\,j_1j_2j_3}(x_1)G^{\rm c}_{j_1i_2}(x_1,x_2)G^{\rm c}_{j_2i_3}(x_1,x_3)G^{\rm c}_{j_3i_4}(x_1,x_4)\nn[2pt]
  &+3Q_{i_1i_2,\,j_1}(x_1,x_2)G^{\rm c}_{j_1i_3i_4}(x_1,x_3,x_4)+3Q_{i_1i_2,\,j_1j_2}(x_1,x_2)G^{\rm c}_{j_1i_3}(x_1,x_3)G^{\rm c}_{j_2i_4}(x_1,x_4)+\mathcal{O}(\varepsilon_q^2\varepsilon^4)\Big]_{\overline{1234}}\,.\label{eq:dtG4_c}\nn[4pt]
\end{align}
\end{subequations}
The diagrammatic representation of Eqs.~\eqref{eq:dtGn_c} is shown in
Fig.~\ref{fig:diagrams}. We find that the leading terms (of order
$\varepsilon_q^2\varepsilon^{n-1}$) are represented by connected tree diagrams. 

\subsection{Arbitrary \texorpdfstring{$n$}{n}}
\label{sec:arbitraryn}

Our derivation can be extended beyond $n\leqslant4$ to higher-order
  $n$-point functions, which are also of interest from the experimental
  point of view \cite{STAR:2022vlo}. In this subsection we discuss
  generalization of 
Eqs.~(\ref{eq:Gn-Gn_c-example}) and~(\ref{eq:dtGn_c})
  to multipoint connected functions of
  arbitrary order $n$. In this work we do not need to
  use such general-$n$ equations, but
  these equations could facilitate the extension of our results to
  higher $n$ in future research. They also shed more light on the
  structure of the $n\leqslant4$ equations we {\em do} use.

  To start, we need to express correlators $G$ in
terms of connected correlators $G^{\rm c}$, similarly to
Eq.~(\ref{eq:Gn-Gn_c-example}).  The correlator $G_{i_1\dots i_n}$ is a sum of
all possible nonequivalent products of $G^{\rm c}$ with indices
${i_1\dots i_n}$ divided between the $G^{\rm c}$ factors in all
possible ways. As in Eq.~(\ref{eq:Gn-Gn_c-example}) we can group these terms
into sets within each of which the difference between the terms is a
permutation of the indices; e.g., $G_{i_1i_2}G_{i_3i_4}+G_{i_1i_3}G_{i_2i_4}+G_{i_1i_4}G_{i_2i_3}$ is such a set. Using the permutation average notation
Eq.~(\ref{eq:aveperm}) we can represent each such a set of terms by a
single term. The resulting combinatorial factors will simply count the
number of terms in each such ``equivalent by permutation'' set. We
thus find the generalization of Eq.~(\ref{eq:Gn-Gn_c-example}) in the form
\begin{equation}\label{eq:Gn-Gn_c}
  G_{i_1\dots i_n}=\sum_{k=1}^n\sum_{\{n_1,\dots,n_k\}}\nf G^{\rm c}_{\underbrace{i_1\dots i_{n_1}}_{n_1}}G^{\rm c}_{\underbrace{i_{n_1+1}\dots i_{n_1+n_2}}_{n_2}}\dots G^{\rm c}_{\underbrace{i_{n-n_k+1}\dots i_{n}}_{n_k}}\bigg|_{\overline{1\dots n}}\,,
\end{equation}
where the inner sum is over all ordered sets of integer numbers
$\{n_1,\dots,n_k\}\in \mathbb N^k$,  such that $n_1\leqslant
n_2\leqslant\dots\leqslant n_k$ and $n_1+\dots+n_k=n$. Each set
describes a partition of the $n$ indices $i_1,\dots,i_n$ into $k$ groups
\begin{equation}\label{eq:partitions}
\underbrace{i_1,\dots,i_{n_1}}_{n_1}| \underbrace{i_{n_1+1},\dots,i_{n_1+n_2}}_{n_2}|\dots|\underbrace{i_{n-n_k+1}\dots i_n}_{n_k}\,,
\end{equation}
in a way that each term in the sum in Eq.~(\ref{eq:Gn-Gn_c}) is different.

The factor $\nf$ is the order of the group of
permutations of $n$ indices $i_1,\dots,i_n$ which leaves the product of $G^{\rm c}$'s
in Eq.~(\ref{eq:Gn-Gn_c}) unchanged due to the symmetry of each $G^{\rm
  c}$ with respect to its own indices as well as the
commutativity of the product of $G^{\rm c}$'s themselves.
In other words, the factor counts
how many identical terms of a given type appear in the sum over {\em
  all} $n!$ permutations of $n$ indices:
\begin{equation}
  \label{eq:nf}
  \nf \equiv \frac{n!}{n_1!\dots n_k!\,k_1!\dots k_n!}\,,
\end{equation}
where $k_m$ is the count of times a given integer $m$ appears in the given set
$\{n_1,\dots,n_k\}$; e.g., for $n=7$ set $\{2,2,3\}$ the counts are
$k_1=0$, $k_2=2$ and $k_3=1$.
The denominator in
Eq.~(\ref{eq:nf}) can be viewed as a symmetry factor corresponding to
a diagram with $k$ vertices (factors of $G^{\rm c}$) with
$n_1,\dots, n_k$ equivalent legs (indices) each. In this picture, $k_m$ is the
number of equivalent vertices with $m$ legs.

By definition, $k_m=0$
for $m>n$ and $\sum_m mk_m=n$. For a given $k$ in the outer sum in
Eq.~\eqref{eq:Gn-Gn_c}
all terms must obey $k_1+\dots+k_n=k$. This ensures that all terms in Eq.~\eqref{eq:Gn-Gn_c}
with the same $k$ are of the same order in $\varepsilon$, i.e.,
$\mathcal O(\varepsilon^{n-k})$, according to
Eq.~(\ref{eq:powercounting}).

The inverse of Eq.~(\ref{eq:Gn-Gn_c}) gives connected correlator $G^{\rm
  c}$ in terms of correlators $G$'s:
\begin{equation}\label{eq:Gn_c-Gn}
  G^{\rm c}_{i_1\dots i_n}=\sum_{k=1}^n\sum_{\{n_1,\dots,n_k\}}(-1)^{k-1}(k-1)!\nf G_{\underbrace{i_1\dots i_{n_1}}_{n_1}}G_{\underbrace{i_{n_1+1}\dots i_{n_1+n_2}}_{n_2}}\dots G_{\underbrace{i_{n-n_k+1}\dots i_{n}}_{n_k}}\bigg|_{\overline{1\dots n}}\,.
\end{equation}
This relation can be obtained by noting that the connected correlator generating function, $g_{\rm c}(\mu)=\sum_{n=1}^\infty\frac{1}{n!}G^{\rm c}_n\mu^n$, and the correlator generating function, $g(\mu)=1+\sum_{m=1}^\infty\frac{1}{m!}G_m\mu^m$, are related by $g_{\rm c}(\mu)=\ln g(\mu)$; i.e.,
\begin{equation}
 \sum_{n=1}^\infty\frac{1}{n!}G^{\rm c}_n\mu^n=\ln\left(1+\sum_{m=1}^\infty\frac{1}{m!}G_m\mu^m\right)=\sum_{k=1}^\infty\frac{(-1)^{k-1}}{k}\left(\sum_{m=1}^\infty\frac{1}{m!}G_m\mu^m\right)^k\,,
\end{equation}
where the logarithmic function is expanded in the last equality. Equating the terms of order $\mu^n$ on both sides, and noting that the contributions from $(\sum_m\frac{1}{m!}G_m\mu^m)^k$ to the terms of order $\mu^n$ are simply given by 
\begin{equation}
 \sum_{\{n_1,\dots,n_k\}}\frac{k!}{k_1!\dots k_n!}\frac{1}{n_1!\dots n_k!}G_{n_1}\dots G_{n_k}\mu^n
\end{equation}
where $n_1+\dots+n_k=n$ and $k_1+\dots+k_n=k$, one obtains Eq.~\eqref{eq:Gn_c-Gn} immediately. For the sake of notation simplicity, we suppress the multivariable indices here, for instance, $G_m\mu^m\equiv G_{i_1\dots i_m}\mu_{i_1}\dots\mu_{i_m}$.

\begin{figure}[ht]
  \centering  
  \includegraphics[scale=.58]{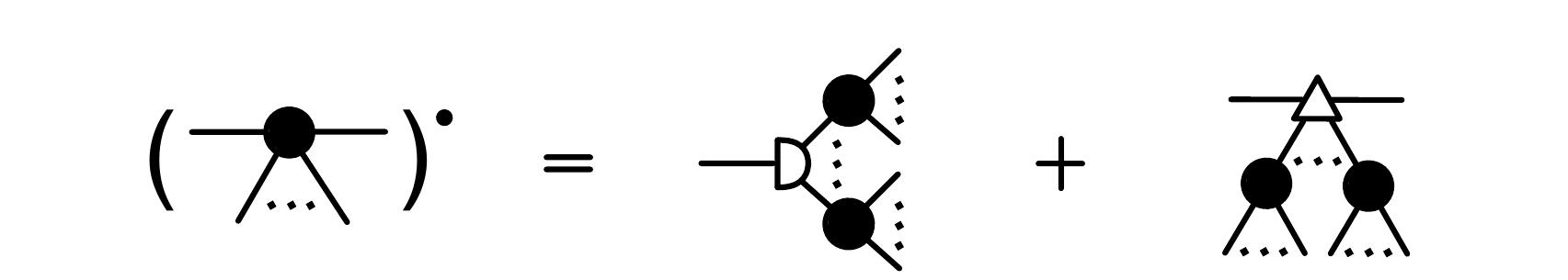}
  \caption{Diagrammatic representation of the evolution equations for generic
    multipoint connected correlation functions. Similarly to
    Fig.~\ref{fig:diagrams}, the first term on the right-hand side of
    Eq.~\eqref{eq:dtGn_c_general} is included
    into the diagram with an open semicircle with two legs. The last two terms are drawn in this figure, with all possible combinatorial arrangements at leading order. See Fig.~\ref{fig:diagrams} for the meaning of the diagram elements, and for the explicit representation of these diagrams when $n=2,3,4$.}
  \label{fig:diagrams-n}
\end{figure}

Using Eqs.~\eqref{eq:dtGn_full}, \eqref{eq:powercounting}, \eqref{eq:Gn-Gn_c}, and \eqref{eq:Gn_c-Gn}, we arrive at the generic equation for $n$-point connected correlation function. At leading order ($\sim\varepsilon_q^2\varepsilon^{n-1}$) it can be obtained by induction from Eqs.~\eqref{eq:dtGn_c} that
\begin{multline}\label{eq:dtGn_c_general}
  \,u\cdot\partial^{(x)} G^{\rm c}_{i_1\dots i_n}(x_1,\dots, x_n)=n\Big[-\left(y_1\cdot\partial u\right)\cdot\frac{\partial}{\partial x_1}G^{\rm c}_{i_1\dots i_n}(x_1,\dots,x_n)\\[2pt]
  +\sum_{k=1}^{n-1}\sum_{\substack{\{n_1,\dots,n_k\}\\n_1+\dots+n_k=n-1}}\nf L_{i_1,\,j_1\dots j_k}(x_1)G^{\rm c}_{j_1\underbrace{i_2\dots}_{n_1}}(x_1,x_2,\dots)\dots G^{\rm c}_{j_k\underbrace{\dots i_n}_{n_k}}(x_1,\dots,x_n)\\[2pt]
  +(n-1)\sum_{k=0}^{n-2}\sum_{\substack{\{n_1,\dots,n_k\}\\n_1+\dots+n_k=n-2}}\nf Q_{i_1i_2,\,j_1\dots j_k}(x_1,x_2)G^{\rm c}_{j_1\underbrace{i_3\dots}_{n_1}}(x_1,x_3,\dots)\dots G^{\rm c}_{j_k\underbrace{\dots i_n}_{n_k}}(x_1,\dots,x_n)\Big]_{\overline{1\dots n}}\,.\\\vspace{0mm}
\end{multline}
The diagrammatic representation of Eq.~\eqref{eq:dtGn_c_general} is sketched in Fig.~\ref{fig:diagrams-n}.

\subsection{Confluent formalism for multipoint equal-time correlators}
\label{sec:confl-form-multi}

An important ingredient for our derivation of the fluctuation evolution equations in {\em relativistic} hydrodynamics is the confluent formalism, which allows us to
covariantly describe fluctuations in the local rest frame of the
fluid. We follow the approach introduced in 
Ref.~\cite{An:2019rhf} for two-point correlators, but generalize it
here to arbitrary $n$-point correlators. Most of the
notations in this section
are similar to the ones introduced in Ref.~\cite{An:2019rhf} and are
summarized in Appendix~\ref{sec:notations} for convenience.

\begin{figure}[ht]
  \centering
  \includegraphics[height=.38\textwidth]{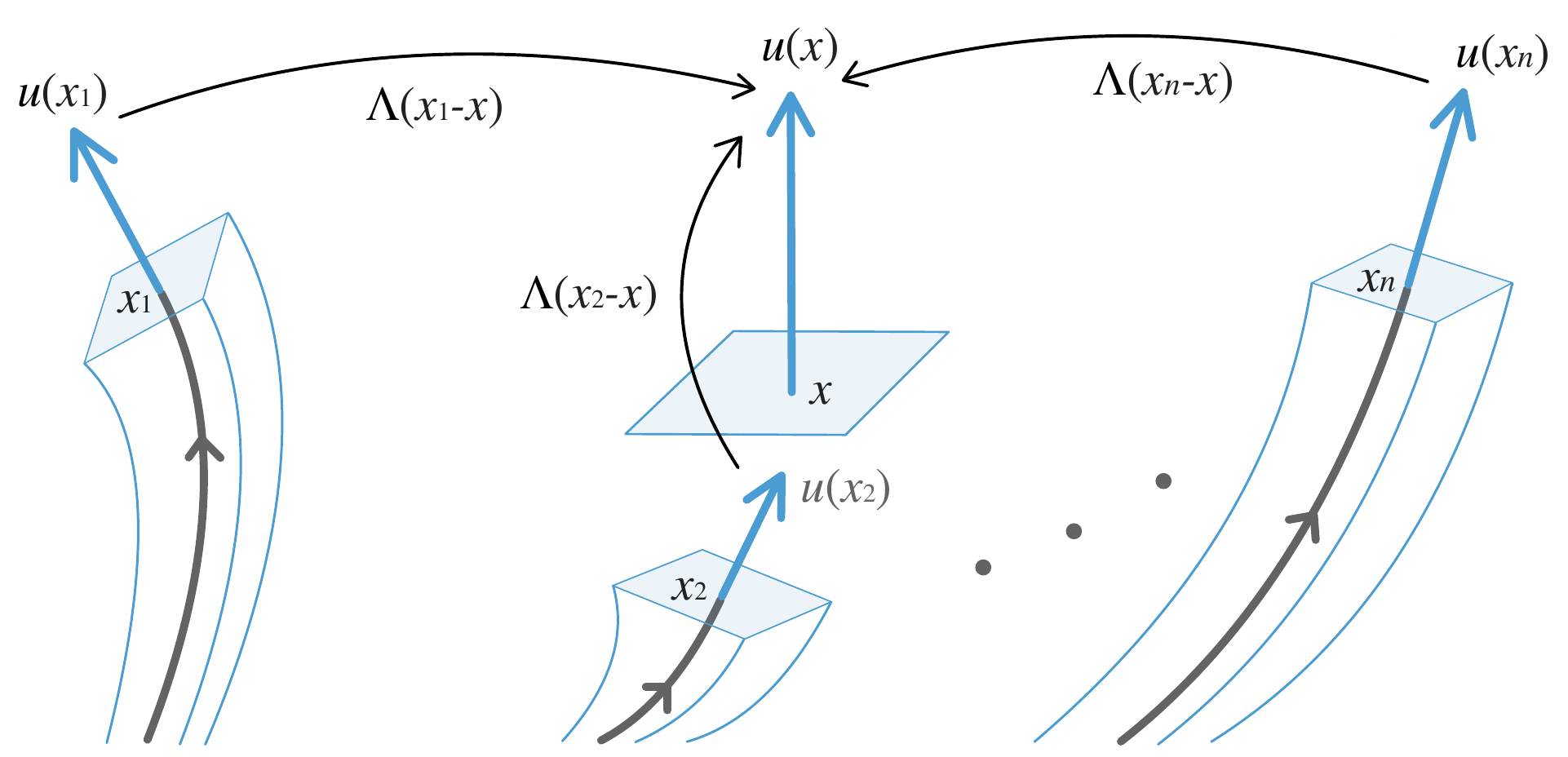}
  \caption{An illustration, in space-time, of the process involved in
    constructing the confluent $n$-point function $\bar G$ from the
    ``raw'' $n$-point function $G$. All $n$ fluctuation fields are
    boosted from their respective local rest frames at $x_i$ to the
    same frame at the midpoint
    $x$.}
  \label{fig:cfc_npt}
\end{figure}

The four-velocities $u$ of the flow in two space-time points $x$ and
$x+\Delta x$ can be related by a Lorentz boost $\Lambda$ which we
define as\footnote{ Our notation, $\Lambda(\Delta x)$, is a short hand for
  $\Lambda(x,\Delta x)$, i.e., boosts in Eq.~(\ref{eq:boost-u})
  depend on two
      points $x$ and $x+\Delta x$. In all equations below the first
      argument is always $x$, and we do not write it explicitly
      to avoid unnecessary clutter.}
\begin{equation}\label{eq:boost-u}
	\Lambda(\Delta x)u(x+\Delta x)=u(x)\,.
      \end{equation}
For infinitesimal $\Delta x$, the four-velocities are different by
infinitesimal amount $\Delta u=(\Delta x\cdot\partial) u$ and the
boost can be represented by the following matrix close to unity
\begin{equation}\label{eq:boost-dx}
  \Lambda(\Delta x)={\mathbb 1}+\Delta\Lambda\,,
  \quad\mbox{where}\quad
  \Delta \Lambda_\mu^{~\nu}(\Delta x)=-u_\mu\Delta u^\nu+u^\nu\Delta u_\mu\,.
\end{equation}
For $n$-point function the fluctuations of the $n$ variables are
evaluated at different points $(x_1, \ldots, x_n)$ with different
local velocities $u(x_i)$. Before comparing or correlating these
local variables it is natural to boost all of them to the same local rest
frame at the midpoint
\begin{equation}
x\equiv
\sum_{i=1}^n\frac{x_i}{n}.\label{eq:x-def}
\end{equation}
If the variables are components of a four-vector
(such as $\delta u^\mu$),
then the boost mixes those components using matrix $\Lambda$ given
explicitly in Eq.~(\ref{eq:boost-dx}). If the variables are scalars
(such as $\delta m$ or $\delta p$)
the boost is trivial (identity), or $\Delta\Lambda=0$. 
To treat these two cases simultaneously in our formalism, we introduce (as
in Ref.~\cite{An:2019rhf}) an object $u^i$ whose components equal to
$u^\mu$ when $i=0,\dots,3$, and zero for all other values of $i$,
corresponding to scalar hydrodynamic variables. 
Then we can express infinitesimal boost from $x+\Delta x$ to $x$, as a matrix
\begin{equation}\label{eq:boost-DeltaLij}
  \Delta\Lambda_{i}^{~j}(\Delta x)= - \Delta x^\mu\bar\omega_{\mu i}^j\,,
\end{equation}
where we defined the confluent connection as in Ref.~\cite{An:2019rhf},
\begin{equation}\label{eq:bar-connection}
\bar\omega_{\mu i}^j = 
  u_{i} \partial_\mu u^{j}-u^{j} \partial_\mu u_{i}\,.
\end{equation}
Boosting all variables into the local rest frame at the midpoint
using $\Lambda(x_i-x)$, we can then define the confluent $n$-point correlator
$\bar G$ and express it in terms of the ``raw'' correlators $G$:
\begin{align}\label{eq:bar-G}
	&\bar G_{i_1\dots i_n}(x_1,\dots,x_n)\equiv\Lambda_{i_1}^{~j_1}(x_1-x)\ldots\Lambda_{i_n}^{~j_n}(x_n-x) G_{j_1\dots j_n}(x_1,\dots,x_n)\nn[2pt]
	\approx\,\,& G_{i_1\dots
                   i_n}(x_1,\dots,x_n)-n\left[
                   (x_1-x)^\mu\bar\omega^{j_1}_{\mu i_1}
                   G_{j_1i_2\dots i_n}(x_1,\dots,x_n)\right]_{\overline{1\dots n}}\,,
\end{align}
where in the second line we have kept only the leading term in the
expansion in powers of $x_i-x$. The idea of the confluent correlator is illustrated in Fig.~\ref{fig:cfc_npt}.

In order to describe the rate of change of a correlation function such
as $\bar G_n$ with respect to the midpoint position
$x$, we want to compare the values of $\bar G_n$ at two sets of
arguments $x_i$ and $x_i'$. The ordinary derivative would correspond to
$x_i'=x_i+\Delta x$ with the same $\Delta x$ for all $i$. A more
natural measure in the context of hydrodynamics, however, would take
into account the change of the four-velocity between the old midpoint $x$
and the new midpoint $x'=x+\Delta x$. Specifically, we want to evaluate the
function at a new set of arguments (points $x_i'$), which are located
relative to the new midpoint $x'$ in exactly the same way (in the
sense to be precisely defined below) as
they were around $x$ {\em in the rest frame} at midpoint. This is
essential if we want to evaluate the rate of change of an {\em
  equal-time} correlator.

That rest
frame defined by the four-velocity $u(x+\Delta x)$, in general, is
different from $u(x)$. The relative position of the points can be
described by four-vectors
\begin{equation}\label{eq:y}
y_i(x)\equiv x_i-x\,.
\end{equation}
In order to preserve the
relative positions in the rest frame at the midpoint we shall define new relative
positions using the same boost as in Eq.~(\ref{eq:boost-u}), i.e.,
$y_i(x+\Delta x) = \Lambda(\Delta x)^{-1}y_i(x)$. This would ensure,
in particular, that the time components of the relative four-vectors $y_i(x)$ in the
rest frame are preserved: $y_i(x)\cdot u(x)=y_i(x+\Delta x)\cdot
u(x+\Delta x)$. Which means that if we define {\em equal-time}
correlator by $y_i(x)\cdot u(x)=0$, the same relation remain true at
the new point: $y_i(x+\Delta x)\cdot u(x+\Delta x)=0$. In other words, if the
points $x_i$ are equal-time in the frame at their midpoint, so are the
points $x'_i$.

Therefore, we define the confluent derivative via the following relation,
where $\Delta x$ is infinitesimal:
\begin{multline}\label{eq:cfd-bar-G-0}
	\Delta x\cdot\cfd\bar G_{i_1\dots i_n}
	\equiv\,\Lambda(\Delta x)_{i_1}^{~j_1}\ldots\Lambda(\Delta x)_{i_n}^{~j_n}\bar G_{j_1\dots j_n}(x_1',\dots,x_n')
   - \bar G_{i_1\dots i_n}(x_1,\dots,x_n)\,,\\[2pt]
    \quad\mbox{where}\quad
    x_i' = x + \Delta x + \Lambda(\Delta x)^{-1}y_i(x)
    \quad\mbox{and}\quad
    x_i = x + y_i(x)
    \,.
  \end{multline}
Note that the variables which are being correlated are also boosted
accordingly to make sure that only their change with respect to the
local rest frame is measured, and not the change of the {\em
  components} of these variables due to the change of
the rest frame itself.
Taking the limit $\Delta x\to 0$ we can write
Eq.~(\ref{eq:cfd-bar-G-0}) in terms of the partial derivatives of $G$
and the boost connection $\bar\omega$:
\begin{align}\label{eq:cfd-barG-2}
  \cfd_\mu\bar G_{i_1\dots i_n}&\equiv n\left[\left(\frac{\partial}{\partial x_1^\mu}-\bar\omega_{\mu\beta}^{\alpha}y_1^\beta\frac{\partial}{\partial x_1^\alpha}\right)\bar G_{i_1\dots i_n}-\bar\omega_{\mu i_1}^{j_1}\bar G_{j_1\dots i_n}\right]_{\overline{1\dots n}}\,.
\end{align}

At this point it is convenient to introduce the derivative with
respect to the midpoint, $x$ which was already defined in Eq.~\eqref{eq:dtGn_full}, as well as the derivatives with respect to
separation vectors $y_i$ {\em at fixed midpoint} $x$:
\begin{align}\label{eq:d/dx-d/dy}
  {\dxisum}\equiv\sum_{i=1}^n\frac{\partial}{\partial x_i}\,,
  \quad \partial^{(y_i)}\equiv\frac{\partial}{\partial x_i}
   - \frac{1}{n}\sum_{j=1}^n\frac{\partial}{\partial x_j}\equiv \frac{\partial}{\partial x_i}
   - \frac{1}{n}\frac{\partial}{\partial x}\,.
\end{align}
Note that $y_i$ variables are not independent, since
$\sum_{i=1}^ny_i=0$, so the derivative $\partial^{(y_i)}$ has
unusual, but simple, properties, e.g., $\partial^{(y_i)} x =0$, $\partial^{(y_i)} y_j =
\delta_{ij} -1/n$.
In terms of such derivatives Eq.~\eqref{eq:cfd-barG-2} reads
\begin{align}\label{eq:cfd-bar-G-diff}
  \cfd_\mu\bar G_{i_1\dots i_n}&\equiv
\dxisum_\mu \bar G_{i_1\dots i_n} -
n\left[ \bar\omega_{\mu\beta}^{\alpha} y_1^\beta\partial^{(y_1)}_\alpha\bar G_{i_1\dots i_n}+\bar\omega_{\mu i_1}^{j_1}\bar G_{j_1\dots i_n}\right]_{\overline{1\dots n}}\,.
\end{align}

\begin{figure}[ht]
  \centering
  \includegraphics[height=.33\textwidth]{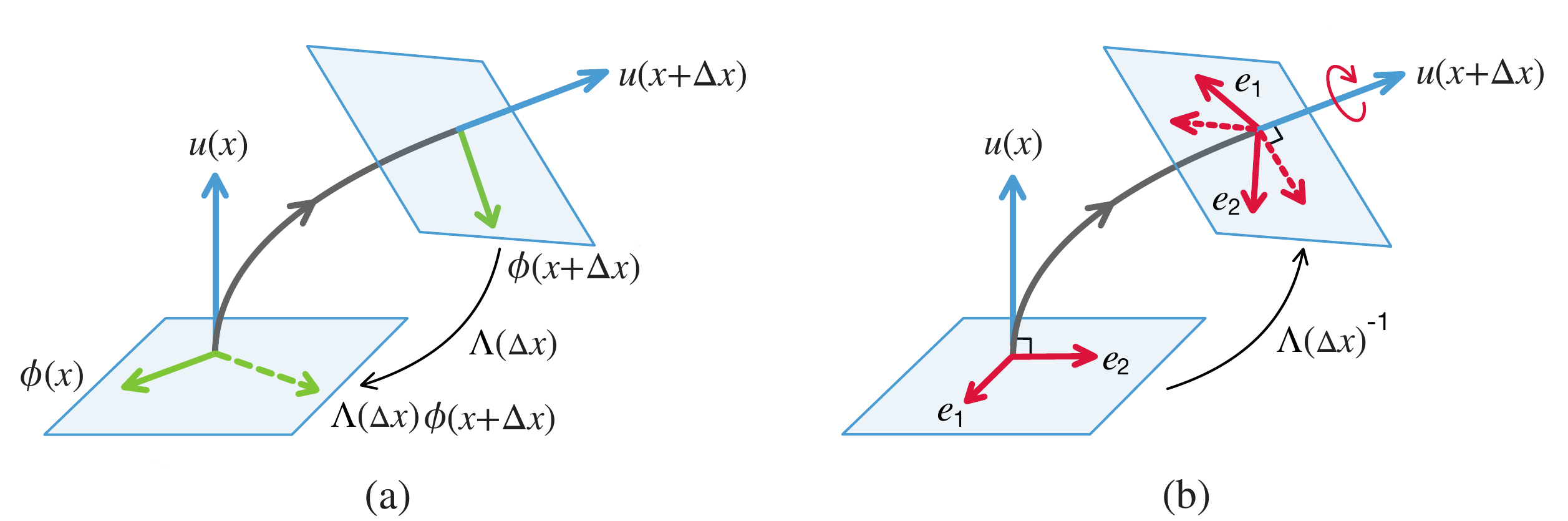}
  \caption{The origin of two connections $\bar\omega$ and $\econ$ in
    the confluent derivative applying to confluent $n$-point
    functions. Panel (a) explains the origin of boost connection
    $\bar\omega$, brought out by boosting the fluctuation variable
    $\phi(x+\Delta x)$ from the local rest frame of $u(x+\Delta x)$ to
    the local rest frame of $u(x)$, i.e.,
    $\Lambda(\Delta x)\phi(x+\Delta x)$, using the same boost that
    takes velocity $u(x+\Delta x)$ to $u(x)$. The action of the boost
    on $\phi$ is determined by the tensor rank of $\phi$ and is illustrated
    in the figure for the case of a four-vector (e.g., velocity
    fluctuation). Panel (b) explains the origin of spin connection
    due to the confluent motion of equal-time surface and the
    arbitrariness of the local triad $e_a$ choice. Under the boost
    from the local rest frame of $u(x)$ to $u(x+\Delta x)$ every
    coordinate vector $y$ in the local equal-time three-hyperplane
    (represented by the blue rectangle), subject to the constraint
    $u(x)\cdot y=0$, must also be boosted simultaneously by
    $\Lambda(\Delta x)^{-1}$. This includes the basis $e_a(x)$ in that
    hyperplane (red vectors, only two dimensions $a=1,2$ are
    shown). Upon the boost the basis vectors map onto $\Lambda(\Delta
    x)^{-1}e_a(x)$ (dashed red vectors) at point $x+\Delta x$ .  The
    additional rotation (indicated by the red circular arrow) in the equal-time plane needed to align the
    boosted basis (dashed red) with the local basis $e_a(x+\Delta x)$ (solid red) gives rise to
    the $\econ$ connection.}
  \label{fig:cfd_npt}
\end{figure}

\begin{figure}[ht]
  \centering
  \includegraphics[height=.33\textwidth]{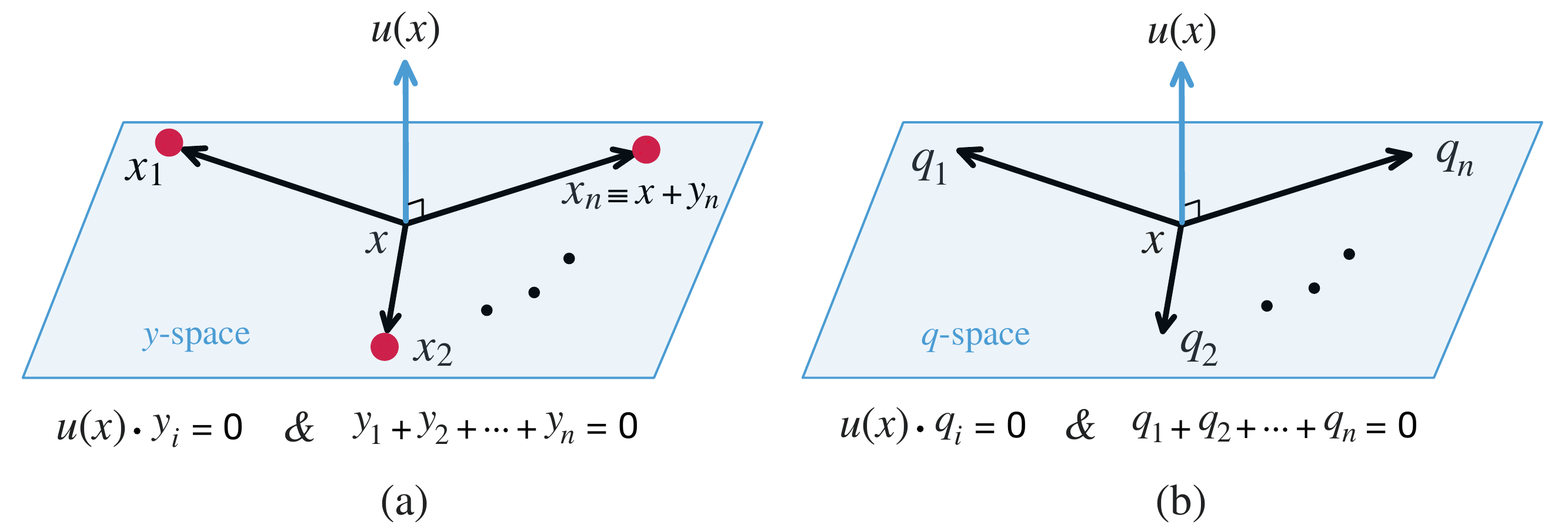}
  \caption{An illustration of the vectors relevant to the Wigner
    transform of equal-time confluent correlators living in
    coordinate $y$-space (shown in panel (a)) to Wigner functions living in
    wave number $\bm q$-space (shown in panel (b)). The wave numbers
    drawn in the figure are four-vectors $e^\mu_aq^a\equiv
    q^\mu$. The nature of the constraint $\sum_i \bm q_i=0$ is
      discussed in the text.
  }
  \label{fig:W_npt}
\end{figure}

Another derivative which will be even more useful when discussing the
Wigner transform is obtained by introducing the local tetrad
consisting of vector $u(x)$ and a triad $e_a(x)$, $a=1,2,3$, thus
expressing vector~$y$ in components: $y(x) =e_a(x)y^a + u(x) y^u$,
where $y^u=-u(x)\cdot y$ is the time component in the local rest frame
at point $x$.  In this paper we shall only consider equal-time
correlators, so we work with $y^u=0$
(cf. Fig.~\ref{fig:W_npt}(a)).
We can then define a derivative where $y^a$ and $y^u=0$ are fixed, i.e.,
\begin{multline}\label{eq:circcfd}
	\Delta x\cdot\dyafxd\bar G_{i_1\dots i_n}
	\equiv\,\bar G_{i_1\dots i_n}
        (x_1',\dots,x_n')
   - \bar G_{i_1\dots i_n}
   (x_1,\dots,x_n)\,,\\[2pt]
    \quad\mbox{where}\quad
    x_i' = x + \Delta x + e_a(x+\Delta x) y^a_i 
    \quad\mbox{and}\quad
    x_i = x + e_a(x) y^a_i
    \,.
  \end{multline}

Taking the limit $\Delta x\to 0$ we find:
\begin{multline}\label{eq:rcfd-bar-G}
    \dyafxd_\mu\bar G_{i_1\dots i_n}
    \equiv n\left[\left(\frac{\partial}{\partial x_1^\mu}
      + \econ_{\mu b}^{a}y_1^b\frac{\partial}{\partial x_1^a}
      + (\partial_\mu u_\lambda) y_1^\lambda
      \left(u\cdot \frac{\partial}{\partial x_1}\right)
    \right)\bar G_{i_1\dots i_n}
    \right]_{\overline{1\dots n}}\\[2pt]
   =\dxisum_\mu \bar G_{i_1\dots i_n}
    +
n\left[ \econ_{\mu b}^{a} y_1^b\partial^{(y_1)}_a
  \bar G_{i_1\dots i_n}
  - \bar\omega_{\mu\lambda}^\nu y_1^\lambda \partial^{(y_1)}_\nu
  \bar G_{i_1\dots i_n}
   \right]_{\overline{1\dots n}}
 \end{multline}
where we introduced
 \begin{align}\label{eq:o-connection}
  \econ_{\mu b}^a\equiv e_\nu^a\partial_\mu e^\nu_b
\end{align}
with $a,b=1,2,3$; the connection associated with the
freedom of choice of the local basis triad $e_a$ at each point
$x$. Using this connection we expressed the derivatives of $e_a$
appearing in Eq.~(\ref{eq:rcfd-bar-G}) as
  \begin{equation}
    \label{eq:de-o}
    \partial_\mu e^\nu_a = (e^\nu_be^b_\lambda
    - u^\nu u_\lambda)\partial_\mu e^\lambda_a
    = e^\nu_b\econ^b_{\mu a}
    + { e^\lambda_a u^\nu (\partial_\mu u_\lambda)}
 = e^\nu_b\econ^b_{\mu a}
     - { e^\lambda_a \bar\omega^\nu_{\mu \lambda } }
  \end{equation}
where we also used Eq.~(\ref{eq:bar-connection}) and $u\cdot y_i=0$. We also
defined the derivative
\begin{equation}
  \label{eq:dya}
  \partial^{(y)}_a\equiv e_a^\mu \partial^{(y)}_\mu\,,
\end{equation}
which is essentially the derivative with respect to three-component
vector $y^a$, which, similarly to $\partial^{(y_i)}_\mu$ defined in Eq.~(\ref{eq:d/dx-d/dy}), is preserving the constraint $\sum_{i=1}^n y_i =0$.

Substituting Eq.~\eqref{eq:rcfd-bar-G} into Eq.~(\ref{eq:cfd-bar-G-diff}), we find
the relationship between the derivatives we defined:
\begin{align}\label{eq:cfd-bar-G-dyafxd}
  \cfd_\mu\bar G_{i_1\dots i_n}&\equiv
\dyafxd_\mu \bar G_{i_1\dots i_n} -
n\left[ \econ_{\mu b}^{a} y_1^b\partial^{(y_1)}_a\bar G_{i_1\dots i_n}+\bar\omega_{\mu i_1}^{j_1}\bar G_{j_1\dots i_n}\right]_{\overline{1\dots n}}\,.
\end{align}

The confluent derivative in Eq.~(\ref{eq:cfd-bar-G-dyafxd}) incorporating the two connections defined in
Eqs.~(\ref{eq:bar-connection}) and~(\ref{eq:o-connection}) is illustrated in Fig.~\ref{fig:cfd_npt}.

Considering now a connected correlation function, $G^\tc(x_1,\ldots,x_n)$ (see Eq.~\eqref{eq:Gn-Gn_c-example}), and using the generalized Wigner transform
  introduced in Ref.~\cite{An_2021} we define the Wigner function
\begin{align}\label{eq:W-ya}
  W_n(x;\bm q_1,\dots, \bm q_n)=\bigintsss \left[\prod_{i=1}^n
  d^3y_i^a\,e^{-iq_{ia}y_i^a}\right]\delta^{(3)}\left(\frac{1}{n}\sum_{i=1}^ny_i^a\right)\bar
  G_n^\tc(x+e_a y_1^a,\ldots,x+e_a y_n^a)\,.
\end{align}
where $q^a$ denote components of three-vector ${\bm
    q}=\{q^a\}\in {\mathbb R}^3$ with $a=1,2,3$ which is the
  wave number conjugate to vector $y^a$. To avoid clutter in our notation, we replaced the indices $i_1,\ldots,i_n$,
which are the same on $W_{i_1\ldots i_n}$ and $\bar
G^\tc_{i_1\ldots i_n}$, with a single index $n$.
The inverse transformation of Eq.~\eqref{eq:W-ya} is given by
  \begin{equation}\label{eq:inverse-WT-ya}
    \bar G_n^\tc(x+e_a y_1^a,\ldots,x+e_a y_n^a)=\bigintsss \left[\prod_{i=1}^n
    \frac{d^3\bm q_i}{(2\pi)^3}\,e^{iq_{ia}
        y_i^a}\right]\delta^{(3)}\left(\sum_{i=1}^n
      \frac{\bm q_i}{2\pi}\right) W_n(x;\bm q_1,\dots, \bm q_n)\,.
\end{equation}
Some of the features of this transformation of variables are
  illustrated in Fig.~\ref{fig:W_npt}. In particular we note that, due
  to the constraint $\sum_i y_i=0$ implemented by the delta function
  in Eq.~(\ref{eq:W-ya}), the Wigner function is invariant with
  respect to the shift of all wave numbers $\bm q_i$ by the same
  vector. That means we can constrain the value of $\sum_i \bm q_i$
  without losing any information about the dependence of $W$ on its
  arguments (of course, this is related to the fact that $W_n$ has one
  argument more than $G_n$). The natural choice of the constraint is
  $\sum_i \bm q_i=0$ as in Eq.~(\ref{eq:inverse-WT-ya}). An intuitive
  way to understand this is to think of $W_n$ as an $n$-point
  ``amplitude'' and of $\bm q_i$ as the corresponding ``momenta''
  flowing in. Then the constraint is simply a reflection of
  ``momentum'' conservation.

It is easy to see that the derivative of the Wigner function with
respect to $x$ at fixed $\bm q_i$'s,
  \begin{equation}
    \label{eq:ringcfdW}
  \Delta x\cdot \dyafxd W_n 
  (x;
  \bm q_1,\ldots,\bm q_n) = 
    W_n(x+\Delta x;\bm q_1,\ldots,\bm q_n)
    - W_n(x; \bm q_1,\ldots,\bm q_n)\,,
  \end{equation}
is the Wigner transform of $\dyafxd \bar G^\tc$; i.e., the derivative
$\dyafxd$ commutes with the Wigner transform.

It is also easy to see that the derivative defined in
Eqs.~(\ref{eq:dya}) and (\ref{eq:d/dx-d/dy}) upon Wigner transform becomes
simply the multiplication by the corresponding $iq_a$:
\begin{equation}
  \label{eq:dya=qa}
  \partial_a^{(y_i)}
  \quad\xlongrightarrow{~~\text{W.T.}~~}
   \quad i q_{ia} \,.
\end{equation}

We shall define $\cfd W$ as Wigner transform of $\cfd \bar
  G^\tc$. Using Eq.~(\ref{eq:cfd-bar-G-diff}) it is then easy to show that
  \begin{equation}
    \label{eq:Wrcfd-Wcfd}
    \cfd_\mu W_{i_1\dots i_n} = \dyafxd_\mu W_{i_1\dots i_n}
    + n\left(\econ_{\mu b}^{a}q_{1a}\frac{\partial}{\partial q_{1b}}
      W_{i_1\dots i_n}
- \bar\omega_{\mu i_1}^{\jj } W_{\jj i_2\dots i_n}
    \right)_{\overline{1\dots n}}\,.
  \end{equation}

Now let us turn to the derivation of the evolution equation of
connected Wigner functions. Applying the confluent derivative in Eq.~(\ref{eq:Wrcfd-Wcfd}) along
$u(x)$ to Eq.~(\ref{eq:W-ya}), we express the result in terms of the
``raw'' connected correlation functions $G^\tc$ as follows:
\begin{align}\label{eq:cfd-W_c}
	&u\cdot\cfd W_{i_1\dots i_n}(x;\bm q_1,\dots,\bm q_n)=u\cdot\dyafxd W_{i_1\dots i_n} - nu^\mu\left(\bar\omega_{\mu i_1}^{\jj } W_{\jj i_2\dots i_n} - \econ_{\mu b}^{a}q_{1a}\frac{\partial}{\partial q_{1b}} W_{i_1\dots i_n}\right)_{\overline{1\dots n}} \nn
	=&\bigintsss \left[\prod_{i=1}^n d^3y_i\,e^{-iq_{ia}y_i^a}\right]\,\delta^{(3)}\left(\frac{1}{n}\sum_{i=1}^ny_i\right)\Big\{u\cdot\partial G^{\rm c}_{i_1\dots i_n}\nn[4pt]
  &\qquad-
    n \left[\left(u^\mu\bar\omega^\nu_{\mu\lambda} y_1^\lambda
    \partial_\nu^{(y_1)} \delta_{i_1}^{\jj } + y_1^\mu\bar\omega^{\jj
    }_{\mu i_1}u\cdot\partial + u^\mu\bar\omega_{\mu i_1}^{\jj
    }\right)G^{\rm c}_{\jj i_2\dots i_n}\right]_{\overline{1\dots n}}
    \Big\}\,,
\end{align}
where we used Eqs.~\eqref{eq:bar-G}, \eqref{eq:cfd-bar-G-diff} and \eqref{eq:rcfd-bar-G}.

We then apply the evolution equations~\eqref{eq:dtGn_c} (and, generically, Eq.~\eqref{eq:dtGn_c_general}) for $G_n^{\rm
  c}$ to convert local rest frame time derivatives $u\cdot\partial$ into spatial
derivatives and perform the inverse Wigner transform using Eq.~\eqref{eq:inverse-WT-ya}. As a result, we express the confluent local rest frame time derivative
of the Wigner function on the left-hand side of Eq.~(\ref{eq:cfd-W_c})
in terms of the Wigner functions themselves, thus obtaining a set of
local evolution equations.

The explicit form of the evolution equations for Wigner
functions will depend on the explicit form of multilinear operators $L$ and $Q$ in Eqs.~\eqref{eq:dtGn_c} that need
to be substituted into Eq.~\eqref{eq:cfd-W_c}. Below we shall obtain
this explicit form for an important subset of correlators in hydrodynamics.

\section{Specific entropy fluctuations in relativistic hydrodynamics}
\label{sec:entropy}

\subsection{Stochastic equation for specific entropy fluctuations}

With the generic formalism  established, we
now apply it to a more specific
hydrodynamic framework governed by the local equations
for the conservation
of energy, momentum, and charge:
\begin{subequations}\label{eq:conservation}
\begin{align}
	&\partial_\mu\sth T^{\mu\nu}=0\,,\\
	&\partial_\mu\sth J^\mu=0\,,
\end{align}
\end{subequations}
supplemented by constitutive relations for the stress tensor $T^{\mu\nu}$ and
charge current $J^\mu$ in the Landau frame for the stochastic relativistic
hydrodynamics \cite{Kapusta:2011gt},
\begin{subequations}\label{eq:T-J}
\begin{align}
	\sth T^{\mu\nu} &=\sth\eps\sth u^\mu\sth u^\nu+\sth p\sth\Delta^{\mu\nu}+\sth\Pi^{\mu\nu}+\sth S^{\mu\nu}\,,\\
	\sth J^\mu &=\sth n\sth u^\mu+\sth\nu^\mu+\sth I^\mu\,,
\end{align}
\end{subequations}
where $\eps$ and $n$ are the energy density and charge density respectively, measured in the local rest frame; $u$ is the four-velocity already introduced in Eq.~\eqref{eq:dtpsi}; $\Delta_{\mu\nu}=g_{\mu\nu}+u_\mu u_\nu$ is the transverse projection operator satisfying $\Delta^{\mu\nu}u_\nu=0$; the pressure $p$, appearing as the coefficient of $\Delta^{\mu\nu}$ in the ideal part of the stress tensor, is determined by chosen primary hydrodynamic variables (such as $\eps$ and $n$) through the equation of state. The explicit forms of the dissipative parts, denoted by $\Pi^{\mu\nu}$ and $\nu^\mu$ for stress tensor and charge current respectively, are determined by applying the second law of thermodynamics \cite{Landau:2013fluid}:
\begin{gather}\label{eq:Pi-nu}
  \Pi^{\mu\nu}\equiv
  -2\eta\sigma^{\mu\nu}-\zeta\theta\Delta^{\mu\nu}\,, \qquad
  \nu^\mu\equiv-\lambda\Delta^\mu_{~\nu}\partial^\nu\alpha\,,
\end{gather}
where $\alpha\equiv\beta\mu\equiv{\mu}/{T}$ is the chemical potential
to temperature ratio, and gradients of velocity are decomposed into the traceless and trace parts where
\begin{gather}\label{u-quantities}
	\sigma^{\mu\nu}\equiv\theta^{\mu\nu}-\frac{1}{3}\theta\Delta^{\mu\nu}\,,
        \quad
        \theta^{\mu\nu}\equiv\frac{1}{2}\Delta^{\mu\alpha}\Delta^{\nu\beta}(\partial_\alpha u_\beta+\partial_\beta u_\alpha)\,, \quad
        \theta\equiv\theta^\mu_\mu=\partial\cdot u\,.
\end{gather}
The transport coefficients, denoted by $\eta$, $\zeta$ and $\lambda$ above, are shear viscosity, bulk viscosity and charge conductivity respectively.

Eqs.~\eqref{eq:T-J}, accompanied by the  constitutive
equations~(\ref{eq:T-J}) become stochastic due to the presence of the
microscopic scale noises $\sth I^\mu$ and $\sth S^{\mu\nu}$. The
noises are defined similarly to Eqs.~\eqref{eq:dtpsi} and \eqref{eq:eta-eta}:
  \begin{equation}
    \label{eq:I-noise}
    \sth I^\mu = \sth {H^I}^\mu_a \eta^I_a\,,
    \quad\mbox{where}\quad
    \langle\eta^I_a(x_1)\eta^I_b(x_2)\rangle = 2\delta_{ab}\delta^{(4)}(x_1-x_2)
    \quad\mbox{and}\quad
    {H^I}^\mu_a=\sqrt\lambda e^{\mu}_a\,,
  \end{equation}
  with $a=1,2,3$ since there are three independent noises (random currents) in the
  local rest frame, and 
  $e^\mu_a$ is an arbitrary spatial basis triad in the local rest frame:
  $e_a\cdot u=0$ (as already defined and discussed in
  Ref.~\cite{An:2019rhf} and in the previous section).  As a result
  $({H^I}^\mu_a){H^I}^\nu_a=\lambda\Delta^{\mu\nu}$
  (note that
  the choice of the triad $e_a$ does not matter, since
  $e^\mu_ae^\nu_a =\Delta^{\mu\nu}$).
  We define $\sth S^{\mu\nu}$ accordingly. Since there are six independent noise variables (corresponding to
  random stress tensor in local rest frame), the components of
  $\sth S^{\mu\nu}$ are expressed in terms of a symmetric rank-three
  random matrix $\eta^S_{ab}$, i.e.,
\begin{multline}
  \label{eq:S-noise}
  \sth S^{\mu\nu} =  \sth {H^S}^{\mu\nu}_{ab}\eta^S_{ab}\,,
    \quad\mbox{where}\quad
    \langle\eta^S_{ab}(x_1)\eta^S_{cd}(x_2)\rangle =
    (\delta_{ac}\delta_{bd}+\delta_{ad}\delta_{bc})\delta^{(4)}(x_1-x_2)
    \\[4pt]\quad\mbox{and}\quad
    {H^S}^{\mu\nu}_{ab} = \sqrt{\frac{T\eta}{2}}
    \left(e^\mu_ae^\nu_b+ e^\mu_be^\nu_a - \frac{2}{3}\Delta^{\mu\nu}\delta_{ab}\right)
    + \sqrt{\frac{T\zeta}{3}} \Delta^{\mu\nu}\delta_{ab}\,.
  \end{multline}
  Similarly to Eqs.~(\ref{eq:I-noise}) and~(\ref{eq:S-noise}), isotropy requires that the random current and stress noises are statistically independent: 
\begin{equation}\label{eq:SI-noise}
\begin{gathered}
 \av{\eta^I_{a}(x_1)\eta^{S}_{bc}(x_2)}=0\,.
\end{gathered}
\end{equation}

We now turn our focus to the fluctuations of
specific entropy (i.e., ratio of entropy density $s$ to charge density $n$), $m\equiv s/n$, which is parametrically the slowest and also
the most
significant hydrodynamic mode
near a liquid-gas critical point, as we already discussed in
Section~\ref{sec:intro}, such a focus also underlies the Hydro+ approach in Ref.~\cite{Stephanov:2018hydro+}.

Specifically, we are going to derive deterministic equations of motion for the
correlators of $m$. To achieve this, we first obtain the stochastic
equation  for the evolution of $\sth m$
starting from the conservation equation~\eqref{eq:conservation}.
We find
\begin{align}\label{eq:dtm}
	\sth u\cdot\partial\sth m&=\sth m_n\partial^\mu\sth\lambda\sth\Delta_{\mu\nu}\partial^\nu\sth\alpha+2\sth m_\eps\sth\eta\sth\sigma^{\mu\nu}\sth\sigma_{\mu\nu}+\sth m_\eps\sth\zeta\sth\theta^2+\xi_m\,,
\end{align}
where
\begin{align}\label{eq:m_n-m_e}
	m_n\equiv\left(\frac{\partial m}{\partial n}\right)_{\!\eps}=-\frac{\beta w}{n^2}\,, \qquad m_\eps\equiv\left(\frac{\partial m}{\partial\eps}\right)_{\!n}=\frac{\beta}{n}
\end{align}
are expressed in terms of independent thermodynamic derivatives chosen in the set I given by Table~\ref{tab:thermo-derivative} in Appendix.~\ref{sec:thermodynamics}, and
\begin{align}\label{eq:xi_m}
	\xi_m=-\sth m_n\partial_\mu\sth I^\mu+\sth m_\eps\sth u_\nu\partial_\mu\sth S^{\mu\nu}\,.
\end{align}

Eq.~\eqref{eq:dtm} has a form similar to Eq.~\eqref{eq:dtpsi} with
$\psi=m$. Therefore the formalism of the previous section can be
applied. To obtain the equation corresponding to Eq.~\eqref{eq:dtphi},
we expand Eq.~\eqref{eq:dtm} in the fluctuation field, $\dm$, as well
as in derivatives applying to fluctuation fields, such as
$\partial\dm=\mathcal O(q)$, up to $\mathcal O(q^2)$.
\footnote{We keep in mind that fluctuations, such as $\delta m$, are
  characterized by wave number $q$ which is parametrically larger than
  the wave number $k$ characterizing the background of mean quantities,
  such as $m$. Thus $\partial\dm\sim q\sim \varepsilon_q$, while
  $\partial m\sim k \sim \varepsilon_q^2$. In the following equation the
  viscous terms contribute at order higher than $\varepsilon_q^2$ if we focus
  on the correlators of $m$ only.}
It is sufficient to truncate the
expansion at order $\delta m^3$ to obtain equations for the four-point
connected functions.\footnote{To obtain the evolution equation for an
  $n$-point connected function at leading order, one needs to expand
  in fluctuation fields, Eq.~\eqref{eq:dtphi}, up to order
  $\phi^{n-1}$, according to Eq.~\eqref{eq:dtGn_full}.}

While we are interested in correlators of $\delta m$, the evolution of
$\delta m$ also depends on fluctuations of other hydrodynamic
variables, in particular, on fluctuations of pressure,
$\delta p$.\footnote{We choose $m$ and $p$ as the independent
  variables in order to profit from the fact that the fluctuations of
  $m$ and $p$ are uncorrelated in equilibrium. In addition, $m$ and
  $p$ represent the basis of normal modes in the ideal
  hydrodynamics. This significantly simplifies the 
  correlator evolution
  equations, as we have already observed in Ref.~\cite{An:2019fdc}. The
  simplification is even more significant for non-Gaussian
  fluctuations.} As a result, the evolution of the correlators of
$\delta m$ will depend on correlators of $\delta p$ as
well.\footnote{In the regime we consider, transverse velocity correlators,
  which ordinarily would mix with the correlators of specific entropy
  (as in the equation for two-point correlator derived in
  Ref.~\cite{An:2019fdc}), can be considered relaxed (on a
  parametrically faster time scale) to their equilibrium values. These
  values are zero by isotropy and, therefore, we need not consider
  velocity fluctuations in this regime. Correlators involving pressure
  also relax faster, but their equilibrium values are not zero, as we
  shall discuss in more detail below.}  As we shall see below, to the
leading order in hydrodynamic (gradient) expansion and in the
regime we consider, we only need to include terms linear in $\delta
p$. Correspondingly, if we only write the terms which will contribute
to the correlator evolution equations in the regime we consider (e.g.,
Eqs.~\eqref{eq:dtG_c-m} or \eqref{eq:LW_c-m}), we find
\begin{multline}\label{eq:dtdm}
  u\cdot\partial\dm=
  :u\cdot\partial\sth m:\,
  \equiv~u\cdot\partial\sth m -\av{u\cdot\partial\sth m} = \,\,
  :L_{m,m}\dm + L_{m,p}\dpp
  + \frac{1}{2}L_{m,mm}\dm\dm + L_{m,mp}\dm\dpp\\
  + \frac{1}{6}L_{m,mmm}\dm\dm\dm + \frac{1}{2}L_{m,mmp}\dm\dm\dpp
  +\xi_m:
\end{multline}

The multilinear operators $L_{m,m\dots m}$ serving as expansion
coefficients are more explicitly written as\footnote{Note that the
  last two terms on the first line
  in Eq.~(\ref{eq:L_mm}) are of order $kq$ and $k^2$, respectively,
  and are neglected on the second line,
  since they are parametrically smaller than
  the leading terms of order $q^2$ being kept.}
\begin{subequations}\label{eq:L_mm}
  \begin{multline}
    L_{m,m}\dm\equiv L_{m,m}[\dm]
    =m_n\partial^\mu \lambda
    \Delta_{\mu\nu}\partial^\nu\alpha_m\dm+m_n\partial^\mu
    \lambda_m\dm
    \Delta_{\mu\nu}\partial^\nu\alpha+(m_n)_{,m}\dm\partial^\mu
    \lambda \Delta_{\mu\nu}\partial^\nu\alpha\\[2pt]
    =m_n\lambda\alpha_m
    \Delta_{\mu\nu}\partial^\mu\partial^\nu\dm + \mathcal O(kq)
    \,,
  \end{multline}
  \begin{multline}
    L_{m,p}\dpp\equiv L_{m,p}[\dpp]
    =m_n\partial^\mu \lambda
    \Delta_{\mu\nu}\partial^\nu\alpha_p\dpp+m_n\partial^\mu
    \lambda_p\dpp
    \Delta_{\mu\nu}\partial^\nu\alpha+(m_n)_{,p}\dpp\partial^\mu
    \lambda \Delta_{\mu\nu}\partial^\nu\alpha\\[2pt]
    = m_n \lambda\alpha_p\Delta_{\mu\nu}\partial^\mu\partial^\nu\dpp + \mathcal O(kq)
    \,,
  \end{multline}
  \begin{multline}
    L_{m,mm}\dm\dm\equiv L_{m,mm}[\dm,\dm]\\[2pt]
    =m_n\lambda\alpha_{mm}\Delta_{\mu\nu}\partial^\mu\partial^\nu\dm^2 +2m_n\lambda_m\alpha_m\Delta_{\mu\nu}\partial^\mu\dm\partial^\nu\dm +2(m_n)_{,m}\lambda\alpha_m\Delta_{\mu\nu}\dm\partial^\mu\partial^\nu\dm+ \mathcal O(kq)\,,
  \end{multline}
  \begin{multline}
    L_{m,mp}\dm\dpp\equiv L_{m,mp}[\dm,\dpp]\\[2pt]
    =m_n\lambda\alpha_{mp}\Delta_{\mu\nu}\partial^\mu\partial^\nu\dm\dpp
    +m_n\lambda_m\alpha_p\Delta_{\mu\nu}\partial^\mu\dm\partial^\nu\dpp
    +m_n\lambda_p\alpha_m\Delta_{\mu\nu}\partial^\mu\dpp\partial^\nu\dm
    \\[2pt]
    +(m_n)_{,m}\lambda\alpha_p\Delta_{\mu\nu}\dm\partial^\mu\partial^\nu\dpp+(m_n)_{,p}\lambda\alpha_m\Delta_{\mu\nu}\dpp\partial^\mu\partial^\nu\dm+ \mathcal O(kq)\,,
  \end{multline}
  \begin{multline}
  L_{m,mmm}\dm\dm\dm\equiv L_{m,mmm}[\dm,\dm,\dm]\\[2pt]
  =m_n\partial^\mu\lambda\alpha_{mmm}\Delta_{\mu\nu}\partial^\nu\dm^3+3m_n\lambda_m\alpha_{mm}\Delta_{\mu\nu}\partial^\mu\dm\partial^\nu\dm^2+3(m_n)_{,m}\lambda\alpha_{mm}\Delta_{\mu\nu}\dm\partial^\mu\partial^\nu\dm^2\\[2pt]
  +3m_n\lambda_{mm}\alpha_m\Delta_{\mu\nu}\partial^\mu\dm^2\partial^\nu\dm+6(m_n)_{,m}\lambda_{m}\alpha_m\Delta_{\mu\nu}\dm\partial^\mu\dm\partial^\nu\dm\\[2pt]
  +3(m_n)_{,mm}\lambda\alpha_m\Delta_{\mu\nu}\dm^2\partial^\mu\partial^\nu\dm+ \mathcal O(kq)\,,
  \end{multline}
\end{subequations}
where
\begin{align}\label{eq:alpha_m}
	\alpha_{\underbrace{_{m\dots m}}_{k}\underbrace{_{p\dots p}}_\ell}\equiv\left(\frac{\partial^{\ell}}{\partial p^\ell}\left(\frac{\partial^{(k)}\alpha}{\partial m^k}\right)_p\right)_{m}\,, \qquad \lambda_{\underbrace{_{m\dots m}}_k\underbrace{_{p\dots p}}_\ell}\equiv\left(\frac{\partial^{(\ell)}}{\partial p^\ell}\left(\frac{\partial^{(k)}\lambda}{\partial m^k}\right)_p\right)_m\,.
\end{align}
In Eqs.~\eqref{eq:L_mm} derivative $\partial$ applies to all factors
to the right of it, and, since we only consider fluctuations of $m$, i.e.,
  $i, j_1, \dots, j_n=m$ and $u_{,m\dots m}=0$ in Eq.~\eqref{eq:L}, we
  have $L_{m,m\dots m}=F_{m,m\dots m}$.

The noise introduced in Eq.~\eqref{eq:xi_m} satisfies
\begin{align}\label{eq:Q_mm}
  Q_{mm}(x_1,x_2)&=\frac{1}{2}\av{\xi_m(x_1)\xi_m(x_2)}=\frac{1}{2}\av{(-\sth m_n\partial_\mu\sth I^\mu+\sth m_\eps\sth u_\kappa\partial_\lambda\sth S^{\lambda\kappa})(x_1)(-\sth m_n\partial_\nu\sth I^\nu+\sth m_\eps\sth u_\beta\partial_\alpha\sth S^{\alpha\beta})(x_2)}\nn[2pt]
  &\approx\frac{1}{2}\av{(\sth m_n\partial_\mu\sth I^\mu)(x_1)(\sth m_n\partial_\nu\sth I^\nu)(x_2)}=m_n(x_1)m_n(x_2)\partial_\mu^{(x_1)}\partial_\nu^{(x_2)}\lambda(x_1)\Delta^{\mu\nu}(x_1)\delta^{(3)}(x_1-x_2)\,,\nn
\end{align}
where in the absence of velocity fluctuation $\du$, terms involving
stress noises $S^{\mu\nu}$ are of a higher order in the gradient expansion, i.e., they are at least of order $kq\sim\varepsilon_q^3$ whereas we truncated Eq.~\eqref{eq:Q_mm} at order $ q^2\sim k\sim \varepsilon_q^2 $.

\subsection{Evolution of non-Gaussian correlation functions}
\label{sec:entropy_equations}

The evolution equations for $G^{\rm c}_{m\dots m}$ can be readily obtained by setting external indices $i\text{'s}=m$ and internal indices $j\text{'s}=(m, p)$ in Eq.~\eqref{eq:dtGn_c}:
\begin{subequations}\label{eq:dtG_c-m}
\begin{align}
  u\cdot\partial G^{\rm c}_{mm}(x_1,x_2)
  =&~2\Big[-\left(y_1\cdot\partial u\right)\cdot\frac{\partial}{\partial x_1}G^{\rm c}_{mm}(x_1,x_2)+L_{m,m}(x_1)G^{\rm c}_{mm}(x_1,x_2)+Q_{mm}(x_1,x_2)\Big]_{\overline{12}}\,,\nn
\\
  u\cdot\partial G^{\rm c}_{mmm}(x_1,x_2,x_3)
  =&~3\Big[-\left(y_1\cdot\partial u\right)\cdot\frac{\partial}{\partial x_1}G^{\rm c}_{mmm}(x_1,x_2,x_3)+L_{m,m}(x_1)G^{\rm c}_{mmm}(x_1,x_2,x_3)\nn
  &+L_{m,p}(x_1)G^{\rm c}_{pmm}(x_1,x_2,x_3)+L_{m,mm}(x_1)G^{\rm c}_{mm}(x_1,x_2)G^{\rm c}_{mm}(x_1,x_3)\nn
  &+2Q_{mm,m}(x_1,x_2)G^{\rm c}_{mm}(x_1,x_3)\Big]_{\overline{123}}\,,
\\[4pt]
  u\cdot\partial G^{\rm c}_{mmmm}(x_1,x_2,x_3,x_4)
  =&~4\Big[-\left(y_1\cdot\partial u\right)\cdot\frac{\partial}{\partial x_1}G^{\rm c}_{mmmm}(x_1,x_2,x_3,x_4)+L_{m,m}(x_1)G^{\rm c}_{mmmm}(x_1,x_2,x_3,x_4)\nn
  &+L_{m,p}(x_1)G^{\rm c}_{pmmm}(x_1,x_2,x_3,x_4)+3L_{m,mm}(x_1)G^{\rm c}_{mm}(x_1,x_2)G^{\rm c}_{mmm}(x_1,x_3,x_4)\nn[3pt]
  &+3L_{m,mp}(x_1)G^{\rm c}_{mm}(x_1,x_2)G^{\rm c}_{pmm}(x_1,x_3,x_4)\nn[3pt]
  &+L_{m,mmm}(x_1)G^{\rm c}_{mm}(x_1,x_2)G^{\rm c}_{mm}(x_1,x_3)G^{\rm c}_{mm}(x_1,x_4)\nn[4pt]
  &+3Q_{mm,p}(x_1,x_2)G^{\rm c}_{pmm}(x_1,x_3,x_4)+3Q_{mm,m}(x_1,x_2)G^{\rm c}_{mmm}(x_1,x_3,x_4)
  \nn
  &+3Q_{mm,mm}(x_1,x_2)G^{\rm c}_{mm}(x_1,x_3)G^{\rm c}_{mm}(x_1,x_4)\Big]_{\overline{1234}}\,,
\end{align}
\end{subequations}
where $L_{m,m\dots m}$'s and $Q_{mm}$ are given by Eqs.~\eqref{eq:L_mm} and \eqref{eq:Q_mm} respectively. In deriving Eqs.~\eqref{eq:dtG_c-m}, we have used $G^{\rm c}_{mp}=0$ following Ref.~\cite{An:2019fdc}.

\begin{figure}[ht]
  \centering  
  \includegraphics[scale=.58]{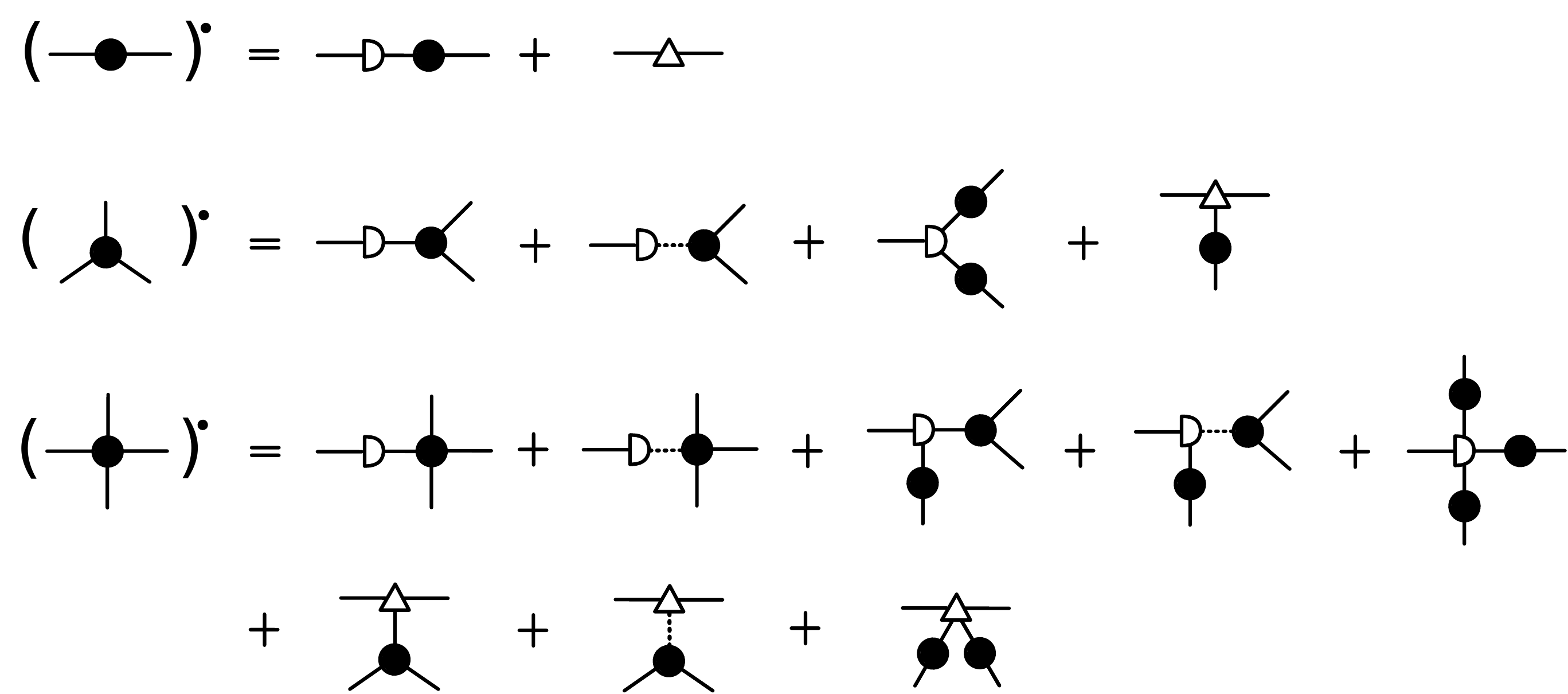}
  \caption{Diagrammatic representation of the evolution
equations~\eqref{eq:LW_c-m}. Similarly to Fig.~\ref{fig:diagrams}, the
solid circles with $k$ legs represent $k$-point Wigner functions $W_k$,
 while the open half-circles represent
$L_{m,\dots}$ (the open half-circles with two solid legs
in the equation for $W_k$ represent $\sim kL_{m,m}+(k-1)\theta$), and
open triangles represent $Q_{mm,\dots}$. The dot on the left-hand side of the
diagrammatic equation represents the Liouville operator in
Eq.~(\ref{eq:LW}). A dashed line leg corresponds to index $p$ replacing
index $m$ in each of the above objects. In the regime we consider,
the Wigner functions involving
pressure fluctuations ($p$ index, or dashed line leg) can be replaced
by their partial equilibrium values given in Eq.~\eqref{eq:Wpeq_mp}
and represented diagrammatically in Fig.~\ref{fig:diagrams-W_peq}. Note that
the diagrams with two-point correlator $W_{mp}$ are not shown, since
this correlator vanishes in the regime we consider.}
  \label{fig:diagrams-W_mm}
\end{figure}

Moreover, for $\psi_i$ being {\em scalars}, Eq.~\eqref{eq:cfd-W_c} is simplified to
\begin{align}\label{eq:cfd-W_c-m}
	u\cdot\cfd W_{i_1\dots i_n}(\bm q_1,\dots,\bm q_n)=&\bigintsss \left[\prod_{i=1}^n d^3y_i\,e^{-iq_{ia}y_i^a}\right]\,\delta^3\left(\frac{1}{n}\sum_{i=1}^ny_i\right)\nn[2pt]
	&\times\left\{u\cdot\partial G^{\rm c}_{i_1\dots i_1}+n\left[(u\cdot\partial u)\cdot y_1 u\cdot\partial^{(y_1)}G^{\rm c}_{i_1\dots i_n}\right]_{\overline{1\dots n}}\right\},
\end{align}
where we suppressed the argument $x$ of $W$. Eq.~\eqref{eq:cfd-W_c-m} applies to our situation when $\psi_i=(m, p)$. Following the procedure discussed at the end of
Sec.~\ref{sec:general}, we substitute Eqs.~\eqref{eq:dtG_c-m} into
Eq.~\eqref{eq:cfd-W_c-m} and perform the inverse Wigner transform
using Eq.~\eqref{eq:inverse-WT-ya}. We then arrive at
\begin{subequations}\label{eq:LW_c-m}
\begin{align}
  \mathcal{L}[W_{mm}(\bm q_1,\bm q_2)]=&~\theta W_{mm}(\bm q_1,\bm q_2)+2\big[L_{m,m}(\bm q_1,-\bm q_1)W_{mm}(\bm q_1,\bm q_2)+Q_{mm}(\bm q_1,\bm q_2)\big]_{\overline{12}}\,,\label{eq:LW2_c-m}\\[4pt]
  \mathcal{L}[W_{mmm}(\bm q_1,\bm q_2,\bm q_3)]=&~2\theta W_{mmm}(\bm q_1,\bm q_2,\bm q_3)+3\big[L_{m,m}(\bm q_1,-\bm q_1)W_{mmm}(\bm q_1,\bm q_2,\bm q_3)\nn[2pt]
  &+L_{m,p}(\bm q_1,-\bm q_1)W_{pmm}(\bm q_1,\bm q_2,\bm q_3)\nn[2pt]
  &+L_{m,mm}(\bm q_1,\bm q_2,\bm q_3)W_{mm}(-\bm q_2,\bm q_2)W_{mm}(-\bm q_3,\bm q_3)\nn[2pt]
  &+2Q_{mm,m}(\bm q_1,\bm q_2,\bm q_3)W_{mm}(-\bm q_3,\bm q_3)\big]_{\overline{123}}\,,\label{eq:LW3_c-m}\\[4pt]
  \mathcal{L}[W_{mmmm}(\bm q_1,\bm q_2,\bm q_3,\bm q_4)]=&~3\theta W_{mmmm}(\bm q_1,\bm q_2,\bm q_3,\bm q_4)+4\big[L_{m,m}(\bm q_1,-\bm q_1)W_{mmmm}(\bm q_1,\bm q_2,\bm q_3,\bm q_4)\nn[2pt]
  &+L_{m,p}(\bm q_1,-\bm q_1)W_{pmmm}(\bm q_1,\bm q_2,\bm q_3,\bm q_4)\nn[2pt]
  &+3L_{m,mm}(\bm q_1,\bm q_2,\bm q_3+\bm q_4)W_{mm}(-\bm q_2,\bm q_2)W_{mmm}(-\bm q_3-\bm q_4,\bm q_3,\bm q_4)\nn[2pt]
  &+3L_{m,mp}(\bm q_1,\bm q_2,\bm q_3+\bm q_4)W_{mm}(-\bm q_2,\bm q_2)W_{pmm}(-\bm q_3-\bm q_4,\bm q_3,\bm q_4)\nn[2pt]
  &+L_{m,mmm}(\bm q_1,\bm q_2,\bm q_3,\bm q_4)W_{mm}(-\bm q_2,\bm q_2)W_{mm}(-\bm q_3,\bm q_3)W_{mm}(-\bm q_4,\bm q_4)\nn[2pt]
  &+3Q_{mm,m}(\bm q_1,\bm q_2,\bm q_3+\bm q_4)W_{mmm}(-\bm q_3-\bm q_4,\bm q_3,\bm q_4)\nn[2pt]
  &+3Q_{mm,p}(\bm q_1,\bm q_2,\bm q_3+\bm q_4)W_{pmm}(-\bm q_3-\bm q_4,\bm q_3,\bm q_4)\nn[2pt]
  &+3Q_{mm,mm}(\bm q_1,\bm q_2,\bm q_3,\bm q_4)W_{mm}(-\bm q_3,\bm q_3)W_{mm}(-\bm q_4,\bm q_4)\big]_{\overline{1234}}\,,
\end{align}
\end{subequations}
where
\begin{equation}\label{eq:LW}
\mathcal{L}[W]\equiv\left(u\cdot\bar\nabla - (\partial_{\nu}u_\mu) e^\mu_a e^\nu_bq_{i}^a\frac{\partial}{\partial q_{ib}}\right) W
\end{equation}
is the Liouville-like operator for the scalar Wigner function \cite{An:2019rhf,An:2019fdc}, and
\begin{equation}\label{eq:L-Q}
\begin{gathered}
L_{m,m}(\bm q_1,\bm q_2)=\gamma_{mm} \bm q_1\cdot \bm q_2\,,  \qquad L_{m,p}(\bm q_1,\bm q_2)=\gamma_{mp} \bm q_1\cdot \bm q_2\,,
	 \\[4pt]
  L_{m,mm}(\bm q_1,\bm q_2,\bm q_3)=-(\gamma_{mm})_{,m} \bm q_1^2+2(\ln m_n)_{,m}\gamma_{mm} \bm q_2\cdot \bm q_3\,,
  \\[4pt]
  L_{m,mp}(\bm q_1,\bm q_2,\bm q_3)=(\gamma_{mm})_{,p}\bm q_1\cdot\bm q_2+(\gamma_{mp})_{,m}\bm q_1\cdot\bm q_3+\left((\ln m_n)_{,m}\gamma_{mp}+(\ln m_n)_{,p}\gamma_{mm}\right)\bm q_2\cdot\bm q_3\,,
  \\[4pt]
	L_{m,mmm}(\bm q_1,\bm q_2,\bm q_3,\bm q_4)=-(\gamma_{mm})_{,mm}\bm q_1^2+2\left((\ln m_n)_{,m}\gamma_{mm}\right)_{,m}(\bm q_2\cdot \bm q_3+\bm q_3\cdot \bm q_4+\bm q_4\cdot \bm q_2)\,,
	 \\[4pt]
	Q_{mm}(\bm q_1,\bm q_2)=-m_n^2\lambda \bm q_1\cdot \bm q_2\,, \qquad Q_{mm,m}(\bm q_1,\bm q_2,\bm q_3)=-(m_n^2\lambda)_{,m}\bm q_1\cdot\bm q_2+(\ln m_n)_{,m}m_n^2\lambda(\bm q_1+\bm q_2)^2\,,
	\\[4pt]
  Q_{mm,p}(\bm q_1,\bm q_2,\bm q_3)=-(m_n^2\lambda)_{,p}\bm q_1\cdot\bm q_2+(\ln m_n)_{,p}m_n^2\lambda(\bm q_1+\bm q_2)^2\,,
  \\[4pt]
  Q_{mp,m}(\bm q_1,\bm q_2,\bm q_3)=-(m_np_n\lambda)_{,m}\bm q_1\cdot\bm q_2-(\ln p_n)_{,m}m_np_n\lambda\bm q_1\cdot\bm q_3-(\ln m_n)_{,m}m_np_n\lambda\bm q_2\cdot\bm q_3\,,
  \\[4pt]
	Q_{mm,mm}(\bm q_1,\bm q_2,\bm q_3,\bm q_4)=-(m_n^2\lambda)_{,mm}\bm q_1\cdot \bm q_2+\left((\ln m_n)_{,m}m_n^2\lambda\right)_{,m}(\bm q_1+\bm q_2)^2-2((\ln m_n)_{,m})^2m_n^2\lambda\bm q_3\cdot\bm q_4\,,
	\\[4pt]
\end{gathered}
\end{equation}
where the coefficients in front of $\bm q_i\cdot\bm q_j$ are thermodynamic derivatives of transport coefficients and thermodynamic quantities. In terms of the independent thermodynamic derivatives chosen from set I in Table~\ref{tab:thermo-derivative}, they are given by
\begin{equation}
\begin{gathered}\label{eq:coeff}
	\gamma_{mm}=m_n\lambda\alpha_m\,, \qquad   \gamma_{mp}=m_n\lambda\alpha_p\,, \qquad (\gamma_{mm})_{,m}=\left(\frac{\lambda_m}{\lambda}+\frac{\alpha_{mm}}{\alpha_m}+\frac{\alpha_p}{\beta m_n}-\frac{2+\alpha_m}{nm_n}\right)\gamma_{mm}\,,\\[4pt]
  (\gamma_{mm})_{,p}=\left(\frac{\lambda_p}{\lambda}+\frac{\alpha_{mp}}{\alpha_m}-\frac{\alpha_p}{nm_n}-\frac{\eps_p}{w}\right)\gamma_{mm}\,,\qquad (\gamma_{mp})_{,m}=\left(\frac{\lambda_m}{\lambda}+\frac{\alpha_{mp}}{\alpha_m}+\frac{\alpha_p}{\beta m_n}-\frac{2+\alpha_m}{nm_n}\right)\gamma_{mp}\,,\\[4pt]
  (\gamma_{mm})_{,mm}=\Bigg[\frac{\lambda_{mm}}{\lambda}+2\left(\frac{\alpha_{mm}}{\alpha_m}+\frac{\alpha_p}{\beta m_n}-\frac{2+\alpha_m}{nm_n}\right)\frac{\lambda_m}{\lambda}+\frac{\alpha_{mmm}}{\alpha_m}\\[4pt]
  +\left(\frac{2\alpha_p}{\beta m_n}-\frac{4+3\alpha_m}{nm_n}\right)\frac{\alpha_{mm}}{\alpha_m}+\frac{\alpha_{mp}}{\beta m_n}-\frac{2\alpha_p}{\beta n m_n^2}+\frac{2+\alpha_m}{n^2m_n^2}\Bigg]\gamma_{mm}\,,\\[4pt]
  (\ln m_n)_{,m}=\frac{\alpha_p}{\beta m_n}-\frac{2+\alpha_m}{nm_n}\,, \qquad (\ln m_n)_{,p}=-\frac{\alpha_p}{nm_n}-\frac{\eps_p}{w}\,,\\[4pt]
  (\ln m_n)_{,mm}=\frac{\alpha_{mp}}{\beta m_n}-\frac{\alpha_{mm}}{nm_n}+\frac{\alpha_p}{\beta nm_n^2}\left(2+2\alpha_m-\frac{n\alpha_p}{\beta}\right)-\frac{(1+\alpha_m)(2+\alpha_m)}{n^2m_n^2}\,,\\[4pt]
  (m_n^2\lambda)_{,m}=\left(\frac{\lambda_m}{\lambda}+\frac{2\alpha_p}{\beta m_n}-\frac{4+2\alpha_m}{nm_n}\right)m_n^2\lambda\,,\qquad (m_n^2\lambda)_{,p}=\left(\frac{\lambda_p}{\lambda}-\frac{2\alpha_p}{nm_n}-\frac{2\eps_p}{w}\right)m_n^2\lambda\,,\\[4pt]
  (m_np_n\lambda)_{,m}=\left[\frac{\lambda_m}{\lambda}+\frac{\alpha_{mp}}{\alpha_p}+\frac{\alpha_p}{\beta m_n\eps_p}\left(\frac{w\alpha_{pp}}{\alpha_p}-\frac{2n\alpha_p}{\beta}+2\eps_p+2\right)-\frac{2}{nm_n}\right]\frac{m_nw\alpha_p}{\beta\eps_p}\lambda\,,\\[4pt]
  p_n=\frac{w\alpha_p}{\beta\eps_p}\,, \qquad (\ln p_n)_{,m}=\frac{\alpha_{mp}}{\alpha_p}+\frac{\alpha_p}{\beta m_n\eps_p}\left(\frac{w\alpha_{pp}}{\alpha_p}-\frac{2n\alpha_p}{\beta}+\eps_p+2\right)+\frac{\alpha_m}{nm_n}\,,\\[4pt]
  (m_n^2\lambda)_{,mm}=\Bigg[\frac{\lambda_{mm}}{\lambda}+4\left(\frac{\alpha_p}{\beta m_n}-\frac{2+\alpha_m}{nm_n}\right)\frac{\lambda_m}{\lambda}-\frac{2\alpha_{mm}}{nm_n}+\frac{2\alpha_{mp}}{\beta m_n}\\[4pt]
  -\frac{2\alpha_p}{\beta nm_n^2}\left(6+2\alpha_m-\frac{n\alpha_p}{\beta}\right)+\frac{2(2+\alpha_m)(3+\alpha_m)}{n^2m_n^2}\Bigg]m_n^2\lambda\,.
\end{gathered}
\end{equation}
The corresponding expressions written in terms of independent thermodynamic derivatives from set II given by Table~\ref{tab:thermo-derivative} are given in Eq.~\eqref{eq:coeff-setII}. It is worthwhile to mention the following relations:
\begin{equation}
  m_n^2\lambda=\frac{\kappa}{n^2}\,, \qquad \alpha_m=-\frac{\beta w}{c_p}\,, \qquad \gamma_{mm}=\frac{\kappa}{c_p}\,,
\end{equation}
where $\kappa$ is the thermal conductivity and $c_p\equiv
Tn\left({\partial m}/{\partial T}\right)_{p}$ is the fixed pressure
heat capacity per unit volume. Eqs.~(\ref{eq:LW_c-m}) are represented
diagrammatically in Fig.~\ref{fig:diagrams-W_mm}.

Note that the arguments $\bm q_i$ in each function
  ($W$, $L$, and $Q$) in Eq.~(\ref{eq:LW_c-m}) sum up to zero,
  according to the constraint $\sum_i \bm q_i=0$ discussed in
  Section~\ref{sec:confl-form-multi} and illustrated in
  Fig.~\ref{fig:W_npt}. Diagrammatically, this can be understood as the
  conservation of ``momenta'' $\bm q_i$ in each element of the
  diagram.

In order to close Eqs.~(\ref{eq:LW_c-m}) we need to supply the
values of $W_{mmp}$ and $W_{mmmp}$, i.e., crosscorrelators involving
fluctuations of pressure. We have not written terms in which the
linear (Gaussian) correlator $W_{mp}$ appears, because this correlator
vanishes upon averaging over the timescales longer than the period of
sound oscillation, as already observed in
Ref.~\cite{An:2019fdc}. The vanishing of the linear correlator $W_{mp}$
significantly simplifies equations~(\ref{eq:LW_c-m}). However, we cannot
drop the nonlinear correlators $W_{mmp}$ and $W_{mmmp}$, since they
do not average to zero. This can be easily seen by computing (cf. Eqs.~(\ref{eq:Weq_setI}) and (\ref{eq:Wmmp-Wmm})) the
equilibrium values of $W_{mmp}$ and $W_{mmmp}$, which are not zero,
unlike the equilibrium value of $W_{mp}$.

Outside of the regime we consider (i.e., at faster timescales) these
nonlinear cross-correlators obey dynamic equations which will be a
part of the full system of equations for fluctuation correlators. In
this paper we do not intend to derive this full system. Instead,  we use the
fact that different correlators relax on parametrically different timescales, with correlators of the specific entropy fluctuations being
the slowest. We use this hierarchy of scales and observe that pressure
fluctuations (upon averaging over the sound oscillations) relax
parametrically faster (i.e., on the timescale of sound
attenuation) than the timescale of the evolution of the specific
entropy fluctuations on which we focus.  This parametric separation of
scales enables the regime we consider, i.e., Hydro+ regime with only one
parametrically slow nonhydrodynamic mode.
Effectively, we can consider the
fluctuations $\delta m$ frozen when we determine the fluctuations of
$\delta p$.

\begin{figure}[ht]
  \centering  
  \includegraphics[scale=.58]{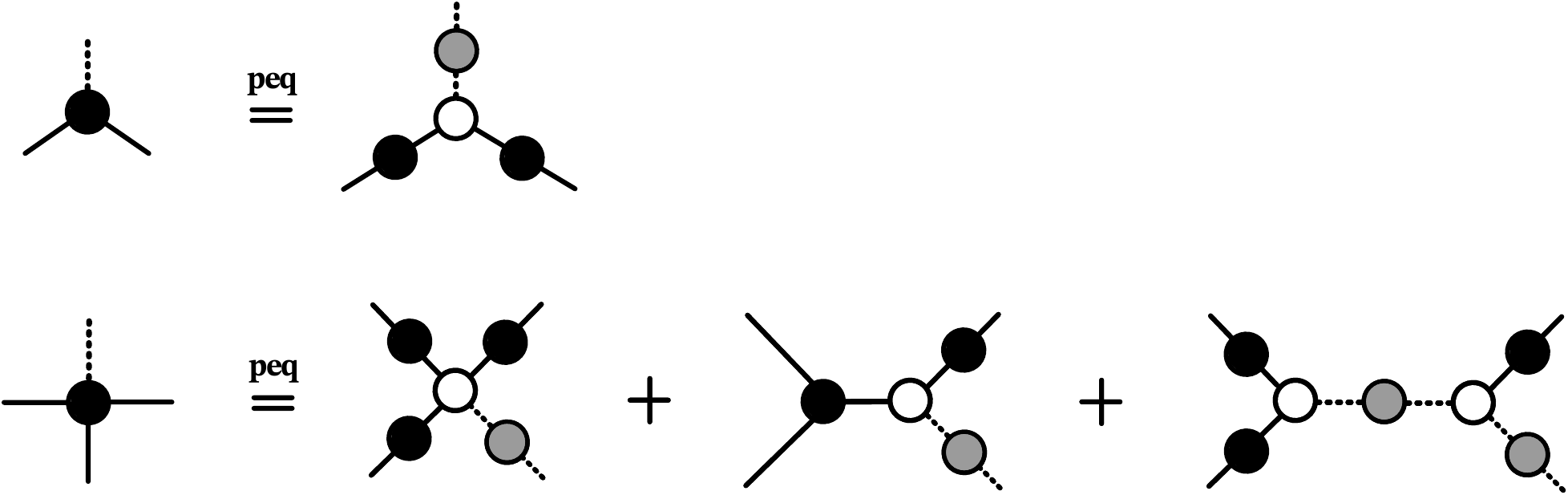}
  \caption{Diagrammatic representation for the
partial equilibrium values of the correlators involving pressure
fluctuation, as given by
Eqs.~(\ref{eq:Wpeq_mp}). An open circle with $k$ legs represents the derivative of entropy
with respect to $k$ variables, e.g., $S_{,mmp}$ for $k=3$, with solid and dashed lines
representing derivatives with respect to $m$ and $p$,
respectively. The solid gray circle with two dashed line legs
represents  $-S^{-1}_{,pp}=W^{\rm eq}_{pp}$, the  equilibrium
value of $W_{pp}$.}
  \label{fig:diagrams-W_peq}
\end{figure}

This means that, while in complete equilibrium, by definition, $\delta p=\delta m=0$, we can also consider {\em partial
  equilibrium} where  $\delta m$ is not zero, i.e., not in equilibrium,
and determine the ``equilibrium'' value of $\delta p$ under these (slowly varying)
conditions. Since fluctuations $\delta p$ relax faster than 
fluctuations $\delta m$, the value of $\delta p$ becomes a function of
$\delta m$, and not an independent variable, in the
regime we consider.
That function is the partial equilibrium value $(\delta p)^{\rm
peq}$, which we derive in Appendix~\ref{sec:part-equil-press}. 
Consequently, the correlators $W_{mmp}$ and $W_{mmmp}$ can be
expressed in terms of the specific entropy correlators as follows:
\begin{equation}\label{eq:Wpeq_mp}
\begin{gathered}
  W_{pmm}^{\rm peq}(\bm q_1, \bm q_2, \bm q_3)
  =-S_{,mmp}S_{,pp}^{-1}W_{mm}(-\bm q_2, \bm q_2)W_{mm}(-\bm q_3, \bm q_3)\,,\\[5pt]
  W_{pmmm}^{\rm peq}(\bm q_1, \bm q_2, \bm q_3, \bm q_4)=-3S_{,mmp}S_{,pp}^{-1}W_{mm}(\bm q_2, -\bm q_2)W_{mmm}(-\bm q_3-\bm q_4, \bm q_3, \bm q_4)\\[2pt]
  -\left(S_{,mmmp}S_{,pp}^{-1}-3S_{,mmp}S_{,mpp}S_{,pp}^{-2}\right)W_{mm}(-\bm q_2, \bm q_2)W_{mm}(-\bm q_3, \bm q_3)W_{mm}(-\bm q_4, \bm q_4)\,.
\end{gathered}
\end{equation}
where $S_{,\dots}$ are derivatives, given by Eqs.~\eqref{eq:S-derivative}, of the entropy functional $S$ that is maximized
in equilibrium, as discussed in Appendix~\ref{sec:thermodynamics}. These equations are represented diagrammatically in Fig.~\ref{fig:diagrams-W_peq}.
Substituting partial equilibrium values for the crosscorrelators
$W_{pmm}$ and $W_{pmmm}$ from Eq.~(\ref{eq:Wpeq_mp}) into Eqs.~(\ref{eq:LW_c-m}) we now obtain a
closed system of equations for correlators of the specific
entropy.\footnote{Interestingly, this substitution generates terms with
  $W_{m\dots m}$ correlators similar to the ones already present in
  Eqs.~(\ref{eq:LW_c-m}). One could say that the effect of the
  pressure fluctuations is to redefine, or renormalize, the
  coefficients $L_{m,m\dots m}$ and $Q_{mm,m\dots m}$.}

In order to check the validity of Eqs.~\eqref{eq:LW_c-m}--(\ref{eq:coeff}) together with
\eqref{eq:Wpeq_mp} we verified that they are solved by the following
space-time and wave number independent values:
\begin{equation}\label{eq:Weq_setI}
\begin{gathered}
  W_{mm}^{\rm eq}=\frac{m_n}{\alpha_m}\,, \quad W_{mp}^{\rm eq}=0\,, \quad W_{pp}^{\rm eq}=\frac{w}{\beta\eps_p}\,,\\[3pt]
  W_{mmm}^{\rm eq}=-\left(\frac{m_n}{\alpha_m}\right)^2\left(\frac{\alpha_{mm}}{\alpha_m}+\frac{4+2\alpha_m}{nm_n}-\frac{3\alpha_p}{\beta m_n}\right)\,, \quad W_{mmp}^{\rm eq}=-\frac{m_n}{\beta\alpha_m}\,,\\[3pt]
  W_{mmmm}^{\rm eq}=\left(\frac{m_n}{\alpha_m}\right)^3\bigg[-\frac{\alpha_{mmm}}{\alpha_m}+3\left(\frac{\alpha_{mm}}{\alpha_m}\right)^2+\frac{2\alpha_{mm}}{nm_n\alpha_m}\left(9+3\alpha_m-\frac{7n\alpha_p}{\beta}\right)+\frac{6\alpha_{mp}}{\beta m_n}+\frac{3\alpha_m\eps_p}{\beta wm_n}\\[2pt]
  ~~~~~~~~+\frac{3(2+\alpha_m)(5+2\alpha_m)}{n^2m_n^2}-\frac{\alpha_p}{\beta n m_n^2}\left(44+16\alpha_m-15\frac{n\alpha_p}{\beta}\right)\bigg]\,,\\[3pt]
  W_{mmmp}^{\rm eq}=\frac{2}{\beta}\left(\frac{m_n}{\alpha_m}\right)^2\left(\frac{\alpha_{mm}}{\alpha_m}+\frac{4+2\alpha_m}{nm_n}-\frac{3\alpha_p}{\beta m_n}\right)\,,
\end{gathered}
\end{equation} 
which, of course, can be obtained from an independent
calculation based on thermodynamics (see
Appendix~\ref{sec:thermodynamics}). This is a
highly nontrivial check of the evolution equations, since it
involves cancellations of many terms, including the
contribution of the pressure fluctuations discussed above.

Eqs.~\eqref{eq:Weq_setI} are expressed in terms of independent
thermodynamic derivatives from set I in
Table~\ref{tab:thermo-derivative}. This is useful for verifying
  that equilibrium correlators in Eqs.~\eqref{eq:Weq_setI} satisfy evolution equations. An alternative representation of Eqs.~\eqref{eq:Weq_setI} using the independent thermodynamic derivatives from set II is given by Eqs.~\eqref{eq:Weq_setII}.

The terms involving $\theta$ reflect trivial expansion effect on the
Wigner function described by the equation $u\cdot\partial W_k  =
(k-1)\theta W_k$. This effect was already discussed in Ref.~\cite{An:2019fdc}
for $k=2$, where it was absorbed by rescaling $W_k$ by a factor of
the density $n$ which also obeys a similar equation due to expansion:
$u\cdot\partial n= -n\theta$. Generalizing this approach to arbitrary $k$
we can eliminate the terms with $\theta$ and simplify the expressions by writing the evolution
equations in terms of rescaled Wigner functions $N$:
\begin{align}\label{eq:N-W}
	N_{\underbrace{_{m\dots m}}_{k}}=n^{k-1}W_{\underbrace{_{m\dots m}}_{k}}\,, \qquad N_{\underbrace{_{pm\dots m}}_{k}}=n^{k-1}W_{\underbrace{_{pm\dots m}}_{k}}\,.
\end{align}
In terms of $N$,
equations~\eqref{eq:LW_c-m} retain exactly the same form, but without
$\theta$ terms (and with extra factor of $n$ for each $Q_{mm,m\dots}$,
which is understandable recalling that $Q_{mm}$ is a two-point Wigner function), i.e.,
\begin{subequations}\label{eq:LN_c-m}
\begin{align}
  \mathcal{L}[N_{mm}(\bm q_1,\bm q_2)]=&~2\big[L_{m,m}(\bm q_1,-\bm q_1)N_{mm}(\bm q_1,\bm q_2)+nQ_{mm}(\bm q_1,\bm q_2)\big]_{\overline{12}}\,,\\[7pt]
  \mathcal{L}[N_{mmm}(\bm q_1,\bm q_2,\bm q_3)]=&~3\big[L_{m,m}(\bm q_1,-\bm q_1)N_{mmm}(\bm q_1,\bm q_2,\bm q_3)\nn[2pt]
  &+L_{m,p}(\bm q_1,-\bm q_1)N_{pmm}(\bm q_1,\bm q_2,\bm q_3)\nn[2pt]
  &+L_{m,mm}(\bm q_1,\bm q_2,\bm q_3)N_{mm}(-\bm q_2,\bm q_2)N_{mm}(-\bm q_3,\bm q_3)\nn[2pt]
  &+2nQ_{mm,m}(\bm q_1,\bm q_2,\bm q_3)N_{mm}(-\bm q_3,\bm q_3)\big]_{\overline{123}}\,,\\[7pt]
  \mathcal{L}[N_{mmmm}(\bm q_1,\bm q_2,\bm q_3,\bm q_4)]=&~4\big[L_{m,m}(\bm q_1,-\bm q_1)N_{mmmm}(\bm q_1,\bm q_2,\bm q_3,\bm q_4)\nn[2pt]
  &+L_{m,p}(\bm q_1,-\bm q_1)N_{pmmm}(\bm q_1,\bm q_2,\bm q_3,\bm q_4)\nn[2pt]
  &+3L_{m,mm}(\bm q_1,\bm q_2,\bm q_3+\bm q_4)N_{mm}(-\bm q_2,\bm q_2)N_{mmm}(-\bm q_3-\bm q_4,\bm q_3,\bm q_4)\nn[2pt]
  &+3L_{m,mp}(\bm q_1,\bm q_2,\bm q_3+\bm q_4)N_{mm}(-\bm q_2,\bm q_2)N_{pmm}(-\bm q_3-\bm q_4,\bm q_3,\bm q_4)\nn[2pt]
  &+L_{m,mmm}(\bm q_1,\bm q_2,\bm q_3,\bm q_4)N_{mm}(-\bm q_2,\bm q_2)N_{mm}(-\bm q_3,\bm q_3)N_{mm}(-\bm q_4,\bm q_4)\nn[2pt]
  &+3nQ_{mm,m}(\bm q_1,\bm q_2,\bm q_3+\bm q_4)N_{mmm}(-\bm q_3-\bm q_4,\bm q_3,\bm q_4)\nn[2pt]
  &+3nQ_{mm,p}(\bm q_1,\bm q_2,\bm q_3+\bm q_4)N_{pmm}(-\bm q_3-\bm q_4,\bm q_3,\bm q_4)\nn[2pt]
  &+3nQ_{mm,mm}(\bm q_1,\bm q_2,\bm q_3,\bm q_4)N_{mm}(-\bm q_3,\bm q_3)N_{mm}(-\bm q_4,\bm q_4)\big]_{\overline{1234}}\,.
\end{align}
\end{subequations}
The equilibrium solutions for Eqs.~\eqref{eq:LN_c-m} can be obtained accordingly, using Eqs.~\eqref{eq:Weq_setI} (or Eqs.~\eqref{eq:Weq_setII}) and \eqref{eq:N-W}.

It is instructive to compare Eqs.~(\ref{eq:LW_c-m}) or~(\ref{eq:LN_c-m}) for fluctuations
of $m$  to similar
evolution equations for the fluctuations of $n$ at fixed $T$ in the charge diffusion problem derived in
Ref.~\cite{An_2021}. The following map between the two problems was conjectured
in Ref.~\cite{An_2021} (this substitution correctly reproduces the equation for
the two-point correlator in Eq.~(\ref{eq:LW2_c-m}), already known from Ref.~\cite{An:2019fdc}):
\begin{equation}
  \label{eq:map-m-n}
  n\to  m\,,\qquad
  \gamma \to \frac{\kappa}{c_p}\,, \qquad
  \alpha' \to \frac{n^2}{c_p}\,.
\end{equation}
While the substitution~\eqref{eq:map-m-n} reproduces the terms in
Eqs.~\eqref{eq:LW_c-m} or \eqref{eq:LN_c-m}, the
expressions for the coefficients in these equations given by Eqs.~(\ref{eq:L-Q})
contain terms with derivatives of
$\ln m_n$ which are not
reproduced by the simple map (\ref{eq:map-m-n}). These terms
reflect the fact that the $\bm q=0$ mode of $n$, i.e.,
$\int\! d^3x\, n$, is a constant of motion, while $\int\!  d^3x\, m$
is not (see Eq.~\eqref{eq:dtm}), unless $m$  happens to
be a linear function of $n$, of course, in which case the derivatives of
$\ln m_n$ vanish. Also, the contributions of the pressure
  fluctuations (the terms with $W_{pm\dots m}$) are not captured by
  the replacement in Eq.~(\ref{eq:map-m-n}).\footnote{One could note,
    however, that the {\em leading\/} critical ($\xi\to\infty$) behavior of the
    coefficients in Eqs.~(\ref{eq:L-Q}) 
   {\em is} reproduced correctly by the substitution
  (\ref{eq:map-m-n}).  }

\section{Conclusions and outlook}
\label{sec:conclusions}

We have generalized the fully Lorentz covariant deterministic approach
to relativistic fluctuating hydrodynamics to non-Gaussian
fluctuations. While the full system of equations involving correlators
of all hydrodynamic variables is still a work in progress, here we
demonstrate how this approach allows us to derive the relativistically
covariant equations for the fluctuations of the slowest hydrodynamic
mode, which is specific entropy, or $m=s/n$, in a
fluid with arbitrary relativistic flow. Such fluctuations are of
special significance near a critical point, in particular, the QCD
critical point, for two reasons. First, the equilibrium fluctuations
of $m$ are the largest near the critical point:
$\langle\delta m^2\rangle\sim c_p\sim\xi^{2-\eta}$. In that sense, this is the
``soft'' direction and the source of the most prominent critical
point signatures.  Second, this mode is also the slowest (not
only it is diffusive, but its diffusion coefficient
$\gamma_\lambda=\kappa/c_p$ vanishes at the critical point as $c_p$
diverges) and, thus, it is the furthest from equilibrium. Therefore, the
non-equilibrium dynamics of this mode of fluctuations are the most
consequential from the point of view of predicting the dynamical effects
on the signatures of the QCD critical point.

Similar evolution equations for non-Gaussian correlators in a static
(nonflowing) fluid at given temperature were derived in
Ref.~\cite{An_2021}. In this paper we consider diffusion of the
slowest diffusive mode in a {\em flowing\/} fluid, more relevant for the
description of the dynamics of critical fluctuations in heavy-ion collisions.

Comparing evolution equations~\eqref{eq:LW_c-m} to the results of
Ref.~\cite{An_2021} we observe many similarities, which were already
anticipated in Ref.~\cite{An_2021}. In fact,
equations~\eqref{eq:LW_c-m} can be obtained from those in
Ref.~\cite{An_2021} by a substitution,
Eq.~\eqref{eq:map-m-n},
as conjectured in
Ref.~\cite{An_2021}. However, the terms
containing derivatives of $\ln m_n$ appearing in the coefficients
given by Eqs.~(\ref{eq:L-Q}) cannot be obtained that
way. Their appearance is related to the qualitative difference between
the density $n$, whose space integral is a conserved quantity, and the
specific entropy $m$,
whose space integral (i.e., $\bm q=0$ mode) is generally {\em not\/}
conserved. We have also found nontrivial contributions of pressure
fluctuations to the evolution of the specific entropy correlators due
to non-linearities of the equation of state (i.e., mode coupling).

As a very nontrivial check of the evolution
equations~\eqref{eq:LW_c-m} and \eqref{eq:L-Q}, we verified that the
equilibrium (thermodynamic) correlators in Eq.~(\ref{eq:Weq_setI})
solve the evolution equations. This requires multiple cancellations
which cannot be achieved without the terms with derivatives of
$\ln m_n$ as well as the contributions of pressure fluctuations.

As we pointed out in the Introduction, our equations include only
tree-level contributions (as did the equations derived in Ref.~\cite{An_2021}). It
could potentially be interesting to extend this analysis to one-loop order and
consider ``long-time tail'' effects on the evolution of
fluctuations. The resulting theory would represent the generalization of
Hydro+ \cite{Stephanov:2018hydro+} to non-Gaussian fluctuations.

Although we limited our analysis to the slowest diffusive mode, we
presented our derivation in a sufficiently general form to facilitate
the extension to fluctuations of faster hydrodynamic modes. The next-to-slowest
modes correspond to fluctuations of the transverse velocity of the
fluid. The theory describing fluctuations of all {\em diffusive\/}
hydrodynamic modes (specific entropy and transverse velocity) was
referred to as Hydro++ in Ref.~\cite{An:2019fdc}. Finally, the full set
of hydrodynamic modes includes pressure/longitudinal velocity
fluctuations, i.e., they include sound -- the fastest hydrodynamic
mode. We leave the development of such a full hydrodynamic theory of
fluctuations to further work.

Despite the focus on the slowest hydrodynamic mode, we believe
the equations presented in this paper are valuable for more
realistic simulations of the dynamical evolution of fluctuations in
heavy-ion collisions (as a first step one can consider extending
exploratory Hydro+ calculations, such as in
Refs.~\cite{Rajagopal:2019hydro,Du:2020bxp,Pradeep:2022mkf}, to
non-Gaussian fluctuations). The ingredients required are the same as in
the usual hydrodynamic simulation: equation of state, e.g., as
in Refs.~\cite{Parotto:2018pwx,Karthein:2021nxe,Kapusta:2021oco}, and
kinetic coefficients.

\acknowledgments

This work is supported by the National Science Centre, Poland, under Grants No. 2018/29/B/ST2/02457 and No. 2021/41/B/ST2/02909 (X.A.), the National Science Foundation CAREER Award No. PHY-2143149 (G.B.), and the U.S. Department of Energy, Office of Science, Office of Nuclear Physics, within the framework of the Beam Energy Scan Theory (BEST) Topical Collaboration and Grant No. DEFG0201ER41195 (M.S., H.-U.Y.). X.A. would like to thank the Isaac Newton Institute for Mathematical Sciences, Cambridge, for support and hospitality during the program ``Applicable resurgent asymptotics: towards a universal theory'' supported by EPSRC Grant No. EP/R014604/1.

\appendix

\section{Non-Gaussian correlators in thermodynamics}
\label{sec:thermodynamics}

In the local rest frame, the total entropy subject to the conservation of charge and energy are given by
\begin{equation}\label{eq:S(m,p)}
  S(m,p)=\int_{\bm x} s(m,p)+\bar\alpha n(m,p)-\bar\beta\epsilon(m,p)\,,
\end{equation}
where we choose our two independent thermodynamic variables as the
specific entropy density $m$ and pressure $p$ associated with
conserved quantities $n$ and $\eps$, i.e., $m=m(n,\eps)$,
$p=p(n,\eps)$ and the variable associated with momentum (e.g.,
velocity $u$) is absent due to the fact that we choose the local rest
frame. $\bar\alpha$ and $\bar\beta$ are the local chemical potential
per temperature and inverse of temperature of the heat bath
respectively.

For a system with two independent thermodynamic variables (such as $m$
and $p$), there are $n+1$ independent $n$th-order thermodynamic
derivatives of entropy. It is convenient for calculations to choose a
basis set of independent derivatives for each order to make sure that
cancellations are easier to carry
out. Table~\ref{tab:thermo-derivative} provides two such basis sets
(set I and II) for derivatives up to fourth order. In set I and II we
have primarily chosen the derivatives of $\alpha$ and $\ln m_n$,
respectively. Of course, one can relate the independent thermodynamic
derivatives from set I to those from set II. Taking the second-order
derivative as an example, we have
\begin{equation}
\begin{gathered}  
  \eps_p=2+\alpha_m+nm_n(\ln m_n)_{,m}-w(\ln m_n)_{,p}\,, \qquad \alpha_p=\beta m_n(\ln m_n)_{,m}+\frac{\beta}{n}(2+\alpha_m)\,,
\end{gathered}
\end{equation} 
while $\alpha_m$ and $m_n/\alpha_m$ are simply functions of each other
and the first-order derivative $m_n=-\beta w/n^2$, so we can also treat $\alpha_m$ instead of $m_n/\alpha_m$ as an independent thermodynamic derivative in set II.

  \newcommand\Tstrut{\rule{0pt}{5.5ex}}     
     \newcommand\Bstrut{\rule[-4ex]{0pt}{0pt}} 
     \setlength{\arrayrulewidth}{0.3mm}
     \setlength{\tabcolsep}{5pt}
    \begin{table}[htb]
    \begin{tabular}{ c | c | c }
    \hline \hline
            order & independent derivatives (set I) & independent derivatives (set II)   \Tstrut\Bstrut\\ 
            \hline
            2nd & $\eps_p,\, \alpha_m,\, \alpha_p$ & $\frac{m_n}{\alpha_m},\, (\ln m_n)_{,m},\, (\ln m_n)_{,p}$ \Tstrut\Bstrut\\
            \hline
            3rd & $ \eps_{pp},\, \alpha_{mm},\, \alpha_{mp},\, \alpha_{pp}$ & $\left(\ln\frac{m_n}{\alpha_m}\right)_{,m},\, (\ln m_n)_{,mm},\, (\ln m_n)_{,mp},\, (\ln m_n)_{,pp}$ \Tstrut\Bstrut\\ 
            \hline
            4th &  $\eps_{ppp},\, \alpha_{mmm},\, \alpha_{mmp},\, \alpha_{mpp},\, \alpha_{ppp}$ & $\left(\ln\frac{m_n}{\alpha_m}\right)_{,mm},\, (\ln m_n)_{,mmm},\, (\ln m_n)_{,mmp},\, (\ln m_n)_{,mpp},\, (\ln m_n)_{,ppp}$  \Tstrut\Bstrut\\
    \hline \hline
    \end{tabular}
    \caption{\label{tab:thermo-derivative}Two sets of independent thermodynamic derivatives chosen in this work. The derivative of quantity $X$ with respect to $m$ (or $p$) while keeping $p$ (or $m$) fixed is denoted by $X_m$ (or $X_p$).}
    \end{table}

As a consequence, all thermodynamic derivatives can be expressed
solely in terms of the independent ones such as those listed in
Table~\ref{tab:thermo-derivative}. In terms of the independent derivatives from set I given by Table~\ref{tab:thermo-derivative}, useful expressions for some second-, third-, and fourth-order thermodynamic derivatives are presented below:
\begin{equation}\label{eq:thermo-derivative}
\begin{gathered}  
n_m=\frac{1}{m_n}\left(1-\frac{n\alpha_p}{\beta}\right)\,, \quad n_p=\frac{n\eps_p}{w}\,, \quad \eps_m=-\frac{w\alpha_p}{\beta m_n}\,, \quad \beta_m=\frac{n\alpha_m}{w}\,, \quad \beta_p=-\frac{\beta}{w}\left(1-\frac{n\alpha_p}{\beta}\right)\,,\\[4pt]
n_{mm}=\frac{1}{m_n}\left[-\frac{n\alpha_{mp}}{\beta}+\frac{2n\alpha_p^2}{\beta^2m_n}+\frac{2+\alpha_m}{nm_n}\left(1-\frac{2n\alpha_p}{\beta}\right)\right]\,, \quad  n_{pp}=-\frac{n}{w^2}\left(w\eps_{pp}-\eps_p\right)\,, \\[4pt]
\eps_{mm}=-\frac{w\alpha_p}{\beta m_n}\left[\frac{\alpha_{mp}}{\alpha_p}+\frac{2}{n m_n}\left(1+\alpha_m-\frac{n\alpha_p}{\beta}\right)\right]\,, \quad \eps_{mp}=-\frac{\alpha_p}{\beta m_n}\left[\frac{w\alpha_{pp}}{\alpha_p}+2\left(1+\eps_p-\frac{n\alpha_p}{\beta}\right)\right]\,, \\[4pt]
n_{mmm}=\frac{1}{m_n}\Bigg[-\frac{n\alpha_{mmp}}{\beta}+\frac{\alpha_{mm}}{nm_n}\left(1-\frac{2n\alpha_p}{\beta}\right)-\frac{\alpha_{mp}}{\beta m_n}\left(7+4\alpha_m-\frac{6n\alpha_p}{\beta}\right)~~~~~~~~~~~~~~~~~~\\
~~~~~~~~~~~~~~~~+6\left(\frac{\alpha_p}{\beta m_n}\right)^2\left(3+2\alpha_m-\frac{n\alpha_p}{\beta}\right)+\frac{(2+\alpha_m)(3+2\alpha_m)}{n^2m_n^2}\left(1-\frac{3\alpha_pn}{\beta}\right)\Bigg]\,,\\[4pt]
\eps_{mmm}=-\frac{w\alpha_p}{\beta m_n}\left[\frac{\alpha_{mmp}}{\alpha_p}+\frac{2\alpha_{mp}}{nm_n\alpha_p}\left(2+2\alpha_m-\frac{3n\alpha_p}{\beta}\right)+6\left(1+\alpha_m-\frac{\alpha_p}{\beta m_n}\right)^2\right]\,.
\end{gathered}
\end{equation} 

Applying the derivative with respect to $m$ or $p$ to the entropy
given by Eq.~\eqref{eq:S(m,p)}, we obtain
\begin{equation}\label{eq:S-derivative}
\begin{gathered}
  S_{,\,m}=0\,, \quad S_{,\,p}=0\,, \quad S_{,\,mm}=-\frac{\alpha_m}{m_n}\,, \quad S_{,\,mp}=0\,, \quad S_{,\,pp}=-\frac{\beta\eps_p}{w}\,, \quad S_{,\,mmp}=-\frac{\alpha_m\eps_p}{wm_n}\,, \\[3pt]
  S_{,\,mmm}=-\frac{\alpha_m}{m_n}\left(\frac{\alpha_{mm}}{\alpha_m}+\frac{2(2+\alpha_m)}{nm_n}-\frac{3\alpha_p}{\beta m_n}\right),  \quad S_{,\,mpp}=\frac{\alpha_p}{wm_n}\left(\frac{w\alpha_{pp}}{\alpha_p}-\frac{2n\alpha_p}{\beta}+\eps_p+2\right)\,,\\[3pt]
  S_{,\,mmmp}=\frac{\alpha_m}{m_n}\left[\frac{3\alpha_{pp}}{\beta m_n}-\left(\frac{\alpha_{mm}}{\alpha_m}+\frac{2+\alpha_m}{nm_n}-\frac{6\alpha_p}{\beta m_n}\right)\frac{\eps_p}{w}+\frac{6\alpha_p}{\beta wm_n}\left(1-\frac{n\alpha_p}{\beta}\right)\right]\,,\\[3pt]
  S_{,\,\,mmmm}=-\frac{\alpha_m}{m_n}\bigg[\frac{\alpha_{mmm}}{\alpha_m}+\frac{2\alpha_{mm}}{nm_n\alpha_m}\left(3+3\alpha_m-\frac{2n\alpha_p}{\beta}\right)-\frac{6\alpha_{mp}}{\beta m_n}+\frac{3(2+\alpha_m)(3+2\alpha_m)}{n^2m_n^2}\\[2pt]
  -\frac{4\alpha_p}{\beta n m_n^2}\left(7+5\alpha_m-\frac{3n\alpha_p}{\beta}\right)\bigg]\,.
\end{gathered}
\end{equation} 
The above expressions for entropy derivatives are evaluated at the
maximum of the entropy $S$ (i.e., in equilibrium), which corresponds to
$\alpha=\bar\alpha$ and $\beta=\bar\beta$. As in Ref.~\cite{An:2019rhf,An:2019fdc}, the choice of $m$ and $p$ as independent thermodynamic variables
makes the second-order entropy derivative quadratic form diagonal, i.e., $S_{,mp}=0$.

The connected $n$-point correlation functions in thermodynamic equilibrium are given by
\begin{equation}\label{eq:Weq-S}
\begin{gathered}
  W^{\rm eq}_{i_1i_2}=-S^{-1}_{,\,i_1i_2}\,, \quad  W^{\rm eq}_{i_1i_2i_3}=-S^{-1}_{,\,i_1j_1}S^{-1}_{,\,i_2j_2}S^{-1}_{,\,i_3j_3}S_{,\,j_1j_2j_3}\,,\\
  W^{\rm eq}_{i_1i_2i_3i_4}=\bigg[S^{-1}_{,\,i_1j_1}S^{-1}_{,\,i_2j_2}S^{-1}_{,\,i_3j_3}S^{-1}_{,\,i_4j_4}S_{,\,j_1j_2j_3j_4}-3S^{-1}_{,\,i_1j_1}S^{-1}_{,\,i_2j_2}S_{,\,j_1j_2j_3}S^{-1}_{,\,j_3j_4}S_{,\,j_4j_5j_6}S^{-1}_{,\,j_5i_3}S^{-1}_{,\,j_6j_4}\bigg]_{\overline{i_1i_2i_3i_4}}\,,
\end{gathered}
\end{equation} 
and generically
\begin{equation}\label{eq:Weq-recursion}
   W^{\rm eq}_{i_1\dots i_n}= \bigg[W^{\rm eq}_{i_1j_1} W^{\rm eq}_{i_2\dots i_n,\,j_1}\bigg]_{\overline{i_1\dots i_n}}\,,
\end{equation} 
where indices $i_n$'s label the external points while indices $j_n$'s label the internal ones. The relation given by Eq.~\eqref{eq:Weq-recursion} easily follows from the cumulant generating function
$g_{\rm c}(\mu_1,\dots,\mu_n)$ such that $W^{\rm eq}_{i_1\dots i_n}=d^ng_{\rm c}/d\mu_1\dots d\mu_n$. Indeed, using the chain
rule on the $\mu$ derivative in $W^{\rm eq}_{i_1\dots i_n}=\left[dW^{\rm eq}_{i_2\dots i_n}/d\mu_1\right]_{\overline{1\dots n}}$
and the fact that (susceptibility) $d\psi_j/d\mu_i=W^{\rm eq}_{ij}$ we obtain Eq.~(\ref{eq:Weq-recursion}).

\begin{figure}[ht]
  \centering  
  \includegraphics[scale=.58]{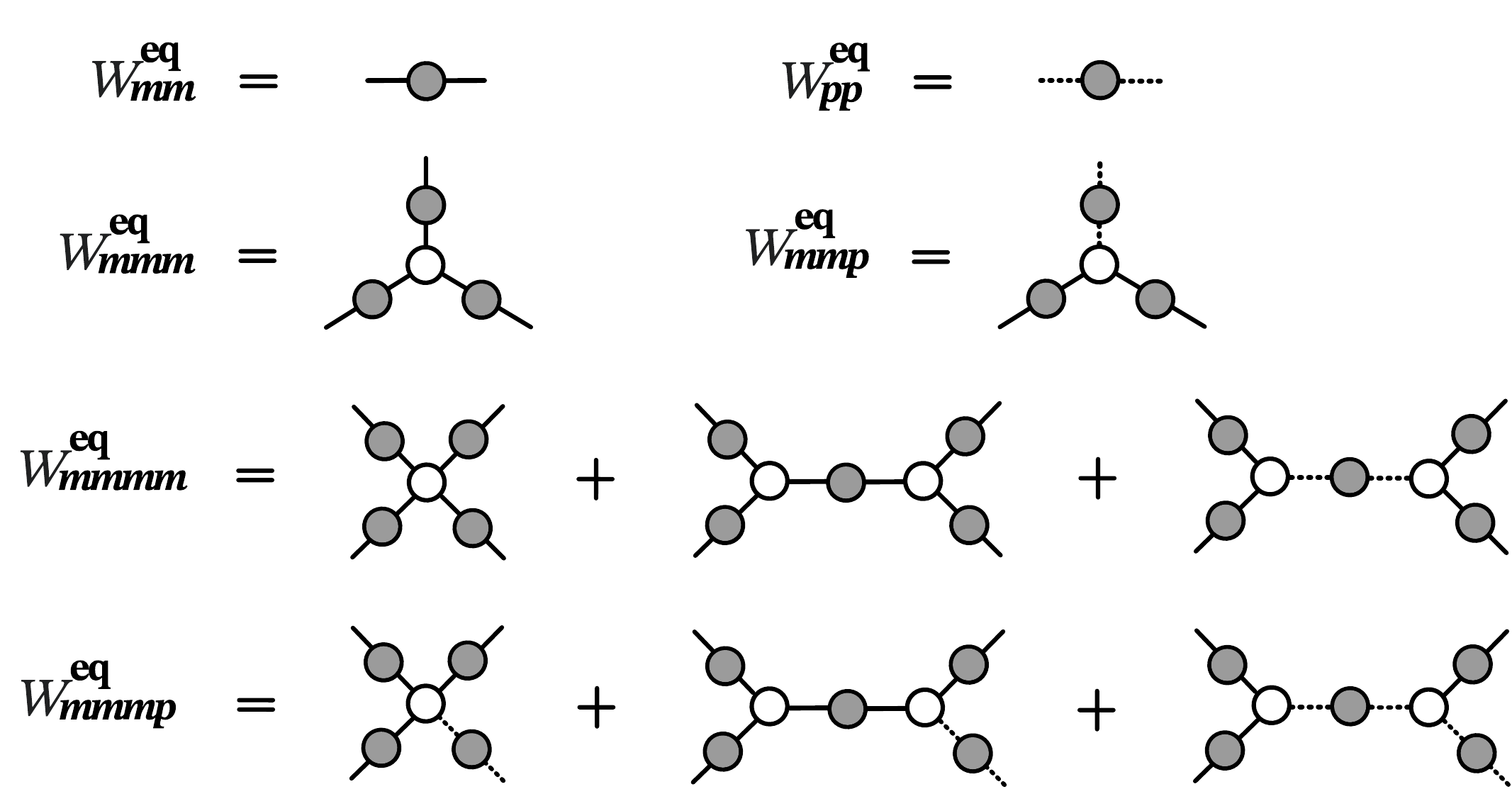}
  \caption{Diagrammatic representation of the equilibrium correlation
    functions in Eq.~\eqref{eq:Weq-S-derivative}. A solid (dashed) line
    entering a vertex denotes the $m$ ($p$) index. Open circles represent
    entropy derivatives. The solid gray circles with two legs are
    equilibrium two-point correlators expressed in terms of the second
    derivatives of the entropy: $W^{\rm eq}_{mm}=-S^{-1}_{mm}$ and $W^{\rm
      eq}_{pp}=-S^{-1}_{pp}$. Note the absence of the cross-correlator
    $W^{\rm eq}_{mp}=0$, which simplifies the calculation.}
  \label{fig:diagrams-Weq}
\end{figure}

In the discussion presented in Sec.~\ref{sec:entropy}, the indices are chosen as $i\text{'s}=m$ and $j\text{'s}=(m,p)$, thus the connected $n$-point functions are (see Fig.~\ref{fig:diagrams-Weq} for diagrammatic representation)
\begin{equation}\label{eq:Weq-S-derivative}
\begin{gathered}
  W^{\rm eq}_{mm}=-S^{-1}_{,\,mm}\,, \quad W^{\rm eq}_{mp}=0\,, \quad W^{\rm eq}_{pp}=-S^{-1}_{,\,pp}\,, \\[3pt]
  W^{\rm eq}_{mmm}=-S^{-3}_{,\,mm}S_{,\,mmm}\,, \quad W^{\rm eq}_{mmp}=-S^{-2}_{,\,mm}S^{-1}_{pp}S_{,\,mmp}\,,\\[3pt]
  W^{\rm eq}_{mmmm}=S^{-4}_{,\,mm}S_{,\,mmmm}-3S^{-5}_{,\,mm}(S_{,\,mmm})^2-3S^{-4}_{,\,mm}S^{-1}_{,\,pp}(S_{,\,mmp})^2\,, \\[3pt]
  W^{\rm eq}_{mmmp}=S^{-3}_{,\,mm}S^{-1}_{,\,pp}S_{,\,mmmp}-3S^{-4}_{,\,mm}S^{-1}_{,\,pp}S_{,\,mmm}S_{,\,mmp}-3S^{-3}_{,\,mm}S^{-2}_{,\,pp}S_{,\,mmp}S_{,\,mpp}\,,
\end{gathered}
\end{equation} 
where we have used the fact that $S_{,mp}=0$ given in Eq.~\eqref{eq:S-derivative}, which largely simplifies our calculation. Now substituting Eq.~\eqref{eq:S-derivative} into Eq.~\eqref{eq:Weq-S-derivative} one immediately obtains Eq.~\eqref{eq:Weq_setI}.

Interestingly, one can infer the following relations from the above expressions:
\begin{equation}\label{eq:Wmmp-Wmm}
\begin{gathered}
  W_{mmp}^{\rm eq}=-\frac{1}{\beta}W_{mm}^{\rm eq}\,, \quad  W_{mmmp}^{\rm eq}=-\frac{2}{\beta}W_{mmm}^{\rm eq}\,.
\end{gathered}
\end{equation} 
The above relations, similarly to $W_{mp}^{\rm eq}=0$, are consequences of choosing $m$ and $p$ as our independent thermodynamic variables.

\section{Partial equilibrium for pressure fluctuations}
\label{sec:part-equil-press}

We shall now use the equilibrium entropy functional given by
Eq.~(\ref{eq:S(m,p)}) to determine the partial equilibrium value of
$\delta p$. This value, $(\delta p)^{\rm peq}$, maximizes the entropy
functional $S$ under the condition that $\delta m$ is fixed. Thus,
$(\delta p)^{\rm peq}$ is determined by solving $S_{,p}(m+\delta
m,p+(\delta p)^{\rm peq})=0$, were $m$ and $p$ are full equilibrium
values, determined by $S_{,m}(m,p)=S_{,p}(m,p)=0$. As a result we
obtain
\begin{equation}\label{eq:deltap-peq}
 (\delta p)^{\rm peq} =\,-\frac{1}{2}S_{,pp}^{-1}S_{,mmp}\dm^2-\left(\frac{1}{6}S_{,pp}^{-1}S_{,mmmp}-\frac{1}{2}S_{,pp}^{-2}S_{,mmp}S_{,mpp}\right)\dm^3\,,
\end{equation}   
where we truncated the solution to the order we need to calculate the
third and fourth order correlators. Notice, that the absence of a term
linear in $\delta m$ is a consequence of the well-known fact that the
fluctuations of pressure and specific entropy are not
correlated, or $S_{,mp}=0$. This lack of correlation, however, does not persist
beyond the linear order. This observation is important for the non-Gaussian
fluctuation equations we derive in this paper.

Using Eq.~(\ref{eq:deltap-peq}), we can now calculate partial
equilibrium values of the cross-correlators of pressure and specific
entropy, such as
\begin{align}\label{eq:G_peq}
G^{\rm c,peq}_{pmm}(x_1,x_2,x_3)\equiv&\,\av{:\!(\delta p)^{\rm peq}(x_1)\!:\dm(x_2)\dm(x_3)}_{\rm c}=-\frac{1}{2}S_{,pp}^{-1}S_{,mmp}\av{:\!\dm(x_1)^2\!:\dm(x_2)\dm(x_3)}_{\rm c}\nn
=&-S_{,pp}^{-1}S_{,mmp}G^{\rm c}_{mm}(x_1,x_2)G^{\rm c}_{mm}(x_1,x_3)\,,\nn[8pt]
G^{\rm c,peq}_{pmmm}(x_1,x_2,x_3,x_4)\equiv&\,\av{:\!(\delta p)^{\rm peq}(x_1)\!:\dm(x_2)\dm(x_3)\dm(x_4)}_{\rm c}\nn
 =&-\frac{1}{2}S_{,pp}^{-1}S_{,mmp}\av{:\dm(x_1)^2:\dm(x_2)\dm(x_3)\dm(x_4)}_{\rm c}\nn
 &-\left(\frac{1}{6}S_{,pp}^{-1}S_{,mmmp}-\frac{1}{2}S_{,pp}^{-2}S_{,mmp}S_{,mpp}\right)\av{:\!\dm(x_1)^3\!:\dm(x_2)\dm(x_3)\dm(x_4)}_{\rm c}\nn
=&-3S_{,pp}^{-1}S_{,mmp}G^{\rm c}_{mm}(x_1,x_2)G^{\rm
   c}_{mmm}(x_1,x_3,x_4)\nn[2pt]
 &-\left(S_{,pp}^{-1}S_{,mmmp}-3S_{,pp}^{-2}S_{,mmp}S_{,mpp}\right)G^{\rm c}_{mm}(x_1,x_2)G^{\rm c}_{mm}(x_1,x_3)G^{\rm c}_{mm}(x_1,x_4)|_{\overline{x_2x_3x_4}}\,,\nn[1pt]
\end{align}
where in the second equality of each above equation, we
neglected the terms which contribute to higher order in fluctuation
expansion parameter $\varepsilon$. Finally, taking the
$n$-point Wigner transform (\ref{eq:W-ya}), we obtain Eqs.~(\ref{eq:Wpeq_mp}).

\section{Alternative expressions in terms of independent thermodynamic derivatives}
\label{sec:indep-coeff}

We have expressed Eqs.~\eqref{eq:coeff} and \eqref{eq:Weq_setI} in terms of the independent thermodynamic derivatives chosen from set I given by Table~\ref{tab:thermo-derivative}. There are, of course, numerous different choices for the independent thermodynamic derivatives in terms of which our equations can be formulated. In this section we provide alternative expressions for these equations, which are instead expressed in terms of independent thermodynamic derivatives specified in set II, where all independent derivatives of $\ln m_n$ (with respect to $m$ or $p$) are chosen as the independent thermodynamic derivatives.

The presence of the derivatives of $m_n$ reflects the fact that the volume
integral of $m$ is not a constant of motion, in contrast to the volume
integral of $n$. The choice of the independent thermodynamic derivatives given by set II allows us to
quantify such a difference. For example, if $m$ were to be a linear
function of $n$, $m_n$ would be a constant and all thermodynamic
derivatives acting on $m_n$ would vanish, significantly simplifying above equations. In this case, there is only one independent $n$th-order thermodynamic derivative (cf.~Table~\ref{tab:thermo-derivative}), as in the charge diffusion problem studied in Ref.~\cite{An_2021}.

The alternative expressions for Eqs.~\eqref{eq:coeff}, in terms of the independent thermodynamic derivatives from set II, read
\begin{equation}
\begin{gathered}\label{eq:coeff-setII}
  \gamma_{mm}=m_n\lambda\alpha_m\,, \qquad   \gamma_{mp}=\left(\frac{2+\alpha_m}{nm_n}+(\ln m_n)_{,m}\right)\beta m_n^2\lambda\,,\\[4pt]
  (\gamma_{mm})_{,m}=\Bigg((\ln\lambda)_{,m}+2(\ln m_n)_{,m}-\left(\ln\frac{m_n}{\alpha_m}\right)_{,m}\Bigg)\gamma_{mm}\,,\\[4pt]
  (\gamma_{mm})_{,p}=\bigg[(\ln m_n)_{,mm}-\frac{\alpha_m}{nm_n}\left(\ln\frac{m_n}{\alpha_m}\right)_{,m}+\left(\frac{2+\alpha_m}{nm_n}+(\ln m_n)_{,m}\right)(\ln m_n)_{,m}\\[2pt]
  +\frac{\alpha_m}{\beta m_n}\left((\ln\lambda)_{,p}+(\ln m_n)_{,p}\right)+\frac{2+\alpha_m}{n^2m_n^2}\bigg]\beta m_n^2\lambda\,,\\[4pt]
  (\gamma_{mp})_{,m}=\bigg[(\ln m_n)_{,mm}-\frac{\alpha_m}{nm_n}\left(\ln\frac{m_n}{\alpha_m}\right)_{,m}\\[2pt]
  +\left(\frac{2+\alpha_m}{nm_n}+(\ln m_n)_{,m}\right)\left((\ln\lambda)_{,m}+2(\ln m_n)_{,m}\right)+\frac{2+\alpha_m}{n^2m_n^2}\bigg]\beta m_n^2\lambda\,,\\[4pt]
  (\gamma_{mm})_{,mm}=\Bigg[(\ln\lambda)_{,mm}+2(\ln m_n)_{,mm}-\left(\ln\frac{m_n}{\alpha_m}\right)_{,mm}\\[2pt]
  +\left((\ln\lambda)_{,m}+2(\ln m_n)_{,m}-\left(\ln\frac{m_n}{\alpha_m}\right)_{,m}\right)^2\Bigg]\gamma_{mm}\,,\\[6pt]
  (m_n^2\lambda)_{,m}=\left((\ln\lambda)_{,m}+2(\ln m_n)_{,m}\right)m_n^2\lambda\,,\qquad (m_n^2\lambda)_{,p}=\left((\ln\lambda)_{,p}+2(\ln m_n)_{,p}\right)m_n^2\lambda\,,\\[8pt]
  (m_np_n\lambda)_{,m}=\left((\ln\lambda)_{,m}+(\ln m_n)_{,m}+(\ln p_n)_{,m}\right)m_np_n\lambda\,,\\[4pt]
  p_n=\frac{w}{n}-\frac{w}{\beta}(\ln m_n)_{,p}\left((\ln m_n)_{,m}+\frac{n(\ln m_n)_{,p}}{\beta}+\frac{2+\alpha_m}{nm_n}\right)^{-1}\,,\\[4pt]
  (\ln p_n)_{,m}=-\frac{2}{n m_n}
   +\frac{nm_n(\ln m_n)_{,mm}-\alpha_m \left(\ln\frac{m_n}{\alpha_m}\right)_{,m}+\frac{2+\alpha_m}{nm_n}}{nm_n(\ln m_n)_{,m}+2+\alpha
   _m}\\[2pt]
   +\frac{\alpha_m\left(\ln\frac{m_n}{\alpha_m}\right)_{,m}-nm_n(\ln m_n)_{,mm}+w(\ln m_n)_{,mp}-w(\ln m_n)_{,p}\left((\ln m_n)_{,m}+\frac{2+\alpha_m}{nm_n}\right)+(\ln m_n)_{,m}}{nm_n(\ln m_n)_{,m}-w(\ln m_n)_{,p}+2+\alpha_m}\,,\\[4pt]
  (m_n^2\lambda)_{,mm}=\left((\ln\lambda)_{,mm}+2(\ln m_n)_{,mm}+((\ln\lambda)_{,m}+2(\ln m_n)_{,m})^2\right)m_n^2\lambda\,.
\end{gathered}
\end{equation}
Note again that we also treat $\alpha_m$ as an independent second-order thermodynamic derivative as $m_n/\alpha_m$, since they are different by $m_n=-\beta w/n^2$ which is only a first-order thermodynamic derivative.

The alternative expressions for Eqs.~\eqref{eq:Weq_setI} likewise read
\begin{equation}\label{eq:Weq_setII}
\begin{gathered}
  W_{mm}^{\rm eq}=\frac{m_n}{\alpha_m}\,, \quad W_{mp}^{\rm eq}=0\,, \quad W_{pp}^{\rm eq}=-\frac{n}{\beta^2}\left(\frac{2+\alpha_m}{nm_n}+(\ln m_n)_{,m}-\frac{w}{nm_n}(\ln m_n)_{,p}\right)^{-1}\,,\\
   W_{mmm}^{\rm eq}=\left(\frac{m_n}{\alpha_m}\right)^2\left[\left(\ln\frac{m_n}{\alpha_m}\right)_{,m}+2(\ln m_n)_{,m}+\frac{2+\alpha_m}{nm_n}\right]\,, \quad W_{mmp}^{\rm eq}=-\frac{m_n}{\beta\alpha_m}\,,\\
  W_{mmmm}^{\rm eq}=\left(\frac{m_n}{\alpha_m}\right)^3\Bigg[\left(\ln\frac{m_n}{\alpha_m}\right)_{,mm}+2\left(\left(\ln\frac{m_n}{\alpha_m}\right)_{,m}+5(\ln m_n)_{,m}+\frac{5+\alpha_m}{nm_n}\right)\left(\ln\frac{m_n}{\alpha_m}\right)_{,m}\\
  +5(\ln m_n)_{,mm}+9\left((\ln m_n)_{,m}+\frac{2+\alpha_m}{nm_n}\right)(\ln m_n)_{,m}-\frac{3\alpha_m}{\beta m_n}(\ln m_n)_{,p}+\frac{(2+\alpha_m)(7+2\alpha_m)}{n^2m_n^2}\Bigg]\,,\\
  W_{mmmp}^{\rm eq}=-\frac{2}{\beta}\left(\frac{m_n}{\alpha_m}\right)^2\left[\left(\ln\frac{m_n}{\alpha_m}\right)_{,m}+2(\ln m_n)_{,m}+\frac{2+\alpha_m}{nm_n}\right]\,.
\end{gathered}
\end{equation} 

\section{Notations}
\label{sec:notations}

This appendix summarizes the notations introduced in Section \ref{sec:confl-form-multi}.

\begin{list}{}{}
\item
$\cfd_\mu$ -- confluent derivative -- Eqs.~(\ref{eq:cfd-bar-G-0}),~(\ref{eq:Wrcfd-Wcfd});
\item 
$\ucon^\nu_{\lambda\mu}$ -- confluent connection -- Eq.~(\ref{eq:bar-connection});
\item 
$\econ^a_{\mu b}$ -- spin connection for local triad $e^a$ --
Eq.~(\ref{eq:o-connection}), Fig.~\ref{fig:cfd_npt};
\item
$\dyafxd_\mu G_n(x_1,\ldots,x_n)$ -- derivative with respect to
the midpoint $x$ at fixed $y_i^a$
-- Eq.~(\ref{eq:circcfd});
\item 
$\cfd_\mu \GG_n$ -- confluent derivative of $\GG_n$ -- Eq.~(\ref{eq:cfd-bar-G-0});
\item
$ G_n \equiv G_{i_1\dots i_n}$ -- ``raw'' $n$-point correlator -- Eq.~(\ref{eq:Gn_def});
\item
$ G_n^\tc \equiv G_{i_1\dots i_n}^\tc$ -- connected $n$-point
correlator --
Eqs.~(\ref{eq:Gn-Gn_c-example}),~(\ref{eq:Gn-Gn_c}),~(\ref{eq:Gn_c-Gn});
\item
$\GG_n\equiv \GG_{i_1\dots i_n}$ -- confluent correlator --
Eq.~(\ref{eq:bar-G}), Fig.~\ref{fig:cfc_npt};
\item
$W_n\equiv W_{i_1\dots i_n}$ -- Wigner transform of connected
confluent correlator
$\GG_n^\tc$ -- Eq.~(\ref{eq:W-ya});
\item
$\dyafxd_\mu \W_n(x;
  \bm q_1,\ldots,\bm q_n)$ -- partial $x$-derivative at fixed 
$\bm q_i$'s (Wigner transform of $\dyafxd_\mu \GG_n^\tc$) -- Eq.~(\ref{eq:ringcfdW}); 
\item
$\cfd_\mu \W_n$ --  confluent derivative of $W_n$ (Wigner transform of $\cfd_\mu \GG_n^\tc$) -- Eq.~(\ref{eq:Wrcfd-Wcfd}); 
\item
$x$ -- the midpoint space-time vector -- Eq.(\ref{eq:x-def});
\item 
$y_i\equiv x_i-x$ -- the separation
four-vector -- Eq.~(\ref{eq:y});
\item
$y_i^a\equiv e^a\cdot y_i$ -- the components of the separation vector
in the local triad basis $e^a(x)$, $a=1,2,3$.
\end{list}

\bibliographystyle{utphys}
\bibliography{references}

\end{document}